\documentclass[aps,pra,onecolumn,superscriptaddress,amssymb,amsfonts,showpacs]{revtex4}

\usepackage[dvips]{graphics,graphicx}
\usepackage{times}

\newcommand{\cA}{{\cal A}}

\newcommand{\cC}{{\cal C}}

\newcommand{\cE}{{\cal E}}

\newcommand{\cH}{{\cal H}}

\newcommand{\cQ}{{\cal Q}}

\newcommand{\cS}{{\cal S}}

\newcommand{\cmplxs}{{\mathbb C}}
\newcommand{\olxs}{{\mathbb O}}

\begin{document}

\title{ Exploring Noiseless Subsystems via Nuclear Magnetic Resonance }

\author{Evan M. Fortunato} 
\affiliation{ Department of Nuclear Engineering,
Massachusetts Institute of Technology, Cambridge, MA 02139 }
\author{ Lorenza Viola} 
\email[Corresponding author: ]{lviola@lanl.gov} 
\affiliation{ Los Alamos National Laboratory, Mail Stop B256, 
Los Alamos, NM 87545} 
\author{ Marco A. Pravia}
\affiliation{ Department of Nuclear Engineering,
Massachusetts Institute of Technology, Cambridge, MA 02139 }
\author{\mbox{Emanuel Knill}}
\affiliation{ Los Alamos National Laboratory, Mail Stop B256, 
Los Alamos, NM 87545} 
\author{Raymond Laflamme}
\affiliation{ Department of Physics, University of Waterloo, 
Waterloo, ON Canada, N2L 3G1, and \\
Perimeter Institute for Theoretical Physics, 35 King Street N., 
Waterloo, ON Canada, N2J 2W9  }
\author{Timothy F. Havel}
\affiliation{ Department of Nuclear Engineering,
Massachusetts Institute of Technology, Cambridge, MA 02139 }
\author{ David G. Cory }
\affiliation{ Department of Nuclear Engineering,
Massachusetts Institute of Technology, Cambridge, MA 02139 }

\begin{abstract}
Noiseless subsystems offer a general and efficient method for 
protecting quantum information in the presence of noise that has 
symmetry properties. A paradigmatic class of error models displaying 
non-trivial symmetries emerges under collective noise behavior, which
implies a permutationally-invariant interaction between the system and 
the environment. We describe experiments demonstrating the preservation 
of a bit of quantum information encoded in a three qubit noiseless 
subsystem for general collective noise.
A complete set of input states is used to determine the super-operator 
for the implemented one-qubit process and to confirm that the fidelity 
of entanglement is improved for a large, non-commutative set of engineered 
errors.  To date, this is the largest set of error operators that has been 
successfully corrected for by any quantum code.  
\end{abstract}

\maketitle

\section{Introduction}

Quantum information processing (QIP) holds the promise of solving problems
in quantum simulation, quantum computation, and secure communication that 
have no known efficient solution in classical information processing 
\cite{NielsenBook}. While QIP can be, in principle, abstractly characterized 
without reference to the details of a specific implementation, physically 
realizing quantum information and its manipulation is essential to practically 
exploit its unique capabilities \cite{Viola-Qubit}. Real physical 
systems are invariably exposed to environmental noise and decoherence due 
to incomplete isolation from their surroundings, as well as to operational 
errors caused by imperfect manipulations. Thus, physical realizations of QIP 
are confronted with the challenge of achieving noise control during storage 
and processing of quantum information.

Thanks to a series of recent investigations, major progress has been witnessed
on the theory front of reliable QIP. On one side, powerful ``accuracy threshold
theorems'' for fault-tolerant quantum error correction (QEC)
\cite{Shor-FaultTol,Kitaev-Algorithms,Aharonov-FaultTol,Knill-Resilient,PreskillRel}
ensure that, if the integrated effect of the noise per qubit and computational
step remains sufficiently small, arbitrarily accurate QIP is still possible, in 
principle, with reasonable resource overheads. On the other side, alternative 
noise control techniques have become available as options complementing and 
expanding the applicability of conventional QEC. These methods include both 
{\sl passive} error control codes based on decoherence-free subspaces (DFSs) 
\cite{Zanardi-NoiselessCodes,DuanGuo,Lidar-DFS} and noiseless subsystems (NSs) 
\cite{Knill-NS,Viola-DynGen,Defilippo,Zanardi-Stabilize}, and {\sl active} error 
suppression schemes relying on dynamical decoupling 
\cite{Viola-DynDecoup,ZanardiSym0,Viola-Control,Tombesi,Viola-EncodedDecoup,Viola-EulerDec}, 
along with a variety of 
device-dependent schemes for reducing systematic and calibration errors 
\cite{Freeman-CompPulse,Shaka-DualComp,JonesComp,Fortunato-Control}. 

In retrospect, what constitutes the unifying conceptual feature of these 
advances, and what ultimately made them possible, is the realization that 
protecting quantum information against noise need {\sl not} require 
the overall state of the physical device supporting QIP to be perfect 
\cite{Knill-QEC}. 
Defining in what sense quantum information can be 
accurately stored in the noisy state of a physical system leads to think of 
all possible noise control options in terms of the emergence of logical 
subsystems (or ``abstract'' quantum particles) which {\sl are} or 
{\sl can be made} immune to noise \cite{Knill-NS,Viola-DynGen,LAScience1} .
In the former case, the occurrence of a NS directly ties into the existence
of symmetry properties of the natural noise process, and error-free 
information storage is ensured without requiring active intervention
\cite{Knill-NS,Viola-Qubit}.
In the latter case, an external control action is instead necessary 
to appropriately symmetrize the dynamics 
\cite{Viola-DynGen,Zanardi-Stabilize,Wu-Create} or to enforce 
noiselessness by a process that incorporates active recovery operations 
as well \cite{Knill-NS}. The basic intuition remains, nevertheless, unchanged.   

While the implications of the subsystem approach are still being 
investigated, the idea of separating the abstract information-carrying 
degrees of freedom from the implemented qubit degrees of freedom has 
proven useful beyond the original motivation of gaining noise protection. 
In particular, the notion of a subsystem has led to an operational 
prescription for realizing logical qubits in physical systems 
\cite{Viola-Qubit}, and the analysis of the resources required to 
universally control subsystem-encoded qubits has resulted in an 
encoded universality approach \cite{Lidar-UniFaultTol,Viola-DynGen}, 
where the notion of universality may be tailored to the set of 
physically available interactions. More recently, the idea of a NS 
has shown to also underlie topological approaches to QIP \cite{Zanardi-Lloyd}. 

From the experimental point of view, the significance of the NS notion 
has only recently begun to be explored. In particular, the first 
implementation of a non-trivial NS for general collective noise in a 
three-qubit liquid-state nuclear magnetic resonance (NMR) quantum 
information processor was reported by Viola {\it et al.} \cite{Viola-NS}.  
Here, we deepen our earlier investigation in two ways: by presenting an 
expanded description of the theoretical analysis and experimental methodology
underlying \cite{Viola-NS}; by reporting additional experimental results 
which may further shed light on the relevance of NSs within present-day 
quantum information technology.

\section{Theoretical background}
\label{Theory}

\subsection{ Noiseless subsystems: 
From simple examples to the general definition }

We begin by trying to build the intuition underlying the NS idea based on
simple considerations and prototype examples \cite{Viola-Qubit,LAScience1}. 
It is a well-known lesson in physics that the occurrence of symmetries in a 
system generally implies the existence of ``conserved quantities'', and that 
these can be exploited to ease the understanding of the system's behavior. 
In its essence, the NS approach adapts this lesson to the QIP-motivated task 
of achieving protection against noise. This is done by noticing that even 
though the system as a whole may be experiencing errors, some of its 
properties may still remain unaffected by them. 
Thus, if information can be represented in terms of the abstract degrees 
of freedom (DOFs) corresponding to such conserved quantities, noiselessness 
may be ensured in spite of the fact that the errors evolve the 
overall system's state. 

This intuition applies to both classical and quantum information storage. A 
simple classical example arises by considering two bits subject to errors which
either flip both bits with probability $p$, or leave them alone with probability 
$1-p$. This error model has the property that it preserves the parity $P(s)$ 
of a state $s$ of the bits, where $P(s)$ is defined as the sum (mod 2) of the 
bit string $s$. Thus, the two values of $P$ can be used together with the
(non-conserved) value of, say, the first physical bit to label the four 
possible states of the two bits: 
\begin{eqnarray*}
00 & \leftrightarrow &  {\tt 0} \cdot 0 \\
01 & \leftrightarrow &  {\tt 1} \cdot 0 \\
10 & \leftrightarrow &  {\tt 1} \cdot 1 \\
11 & \leftrightarrow &  {\tt 0} \cdot 1 \:.
\end{eqnarray*}

The above table establishes a correspondence between the state space of 
the physical system (left hand side) and the one of a pair of abstract 
subsystems (right hand side). In the resulting representation, the first 
member of the abstract pair (the parity bit) carries the information to 
be protected, while the second (the ``syndrome'' bit) experiences the 
effect of the errors.  In this case, parity provides a {\sl classical NS}, 
and the information resides in this protected DOF.

Many of the above features carry over to the quantum-mechanical case. In 
quantum systems, the presence of symmetries (hence of conserved quantities) 
is associated with the existence of operators that commute with all possible 
errors.  First, consider a simple two-qubit example which corresponds to 
complete depolarization on qubit 2.
This error model is defined by the set of error operators 
$\{ E_a\} =\{ .5 \,\sigma_a^{(2)}, \: a=0,x,y,z \}$, $\sigma_0^{(2)}
=\openone^{(2)}$.  Clearly, {\sl no} joint state of the 
two qubits is protected. However, a one-qubit state can be stored in 
the first physical qubit without being affected by the errors: 
qubit 1 is a ``trivial'' quantum NS. 
Mathematically, this intuition is made precise by observing that, 
because $[E_a, \sigma_u^{(1)}]=0$ for every error and $u=x,y,z$, all 
the expectations of $ \sigma_u^{(1)}$ -- hence the state of qubit 1 -- 
are protected from noise: $\sigma_u^{(1)}, u=x,y,z$, define the observables
of this trivial NS qubit.

In general, knowing the {\sl symmetries} of the error model suffices for 
identifying possible protected DOFs, in a way similar to what happens in 
the above simple example: one first determines the set of operators that 
commute with the possible errors, and then in this set identifies 
appropriate combinations that algebraically behave like the observables 
(the Pauli operators) for abstract qubits \cite{Viola-Qubit}. 
If $S$ is a (finite-dimensional) quantum system interacting with some 
environment $E$, a description of noise on $S$ which suffices for 
discussing error protection and error correction properties can be 
obtained by constructing the appropriate {\sl interaction algebra}
 $\cA$ \cite{Knill-NS}. Let the open-system evolution of $S$ involve 
coupling operators $J_a$, where the $J_a$ are traceless and we 
assume, for the moment, that the internal Hamiltonian $H_S$ of $S$ 
alone can be either set to zero or identified with one of the $J_a$.  
Then $\cA$ contains all the complex linear combinations of arbitrary 
products of the $J_a$ and the identity. 
If $S$ is initialized in the state $\varrho_{in}$, and 
the evolution is depicted in terms of a quantum operation
\cite{KrausBook},
\begin{equation}
\varrho_{in} \mapsto \varrho_{out} = {\cE}(\varrho_{in}) = \sum_a E_a 
\varrho_{in} E_a^\dagger \:, \hspace{5mm}   \sum_a E_a^\dagger  E_a =\openone \:,
\label{qop}
\end{equation}
then $\cA$ collects all the possible errors that the 
coupling to $E$ can induce for arbitrary strength or interaction time. 
The {\sl commutant} $\cA'$ of $\cA$, which collects all operators on $S$ that 
commute with arbitrary errors in $\cA$, is the relevant structure to be used in 
generalizing the symmetry argument given above. By construction, both $\cA$ 
and $\cA'$ are (multiplicative) sub-algebras of the full algebra $\cA_S$ of 
operators on $S$. Within the Hamiltonian framework for the composite system 
$S$, $E$ adopted here, they are also naturally closed under the $\dagger$ 
operation, making some standard results from the representation theory of 
operator algebras directly applicable \cite{BurrowBook,Arveson,Landsman}. 
The symmetry properties of error models corresponding to $\cA$ can be 
visualized by thinking of the largest group of unitary operators 
contained in $\cA'$ as the symmetry group for the problem. 

Suppose that $S$ consists of $n$ physical qubits, in which case the state 
space $\cH \simeq (\cmplxs^2)^{\otimes n}$, and $\cA_S$ can be identified 
with the algebra of complex matrices Mat$_N(\cmplxs)$ acting on 
$\cmplxs^N$, $N=2^n$. Then three possibilities are worth examining: 
\begin{itemize}
\item If $\cA'$ consists only of scalar multiples of the $\openone$, 
no useful symmetries are present. Mathematically, this condition 
is equivalent to the fact that the error algebra $\cA$ acts 
``irreducibly'' on $\cH$, hence $\cA = \cA_S$ by the Schur lemma 
\cite{Knill-NS,BurrowBook}.  Because the error process is contributed 
by all operators on $S$, no protected DOF exists.

\item Suppose instead that $\cA'$ is non-trivial, which implies 
that $\cA$ is a proper sub-algebra of $\cA_S=$ Mat$_N(\cmplxs)$. 
If $\cA \simeq \text{Mat}_m(\cmplxs)$ 
for some integer $m$, then one can show \cite{BurrowBook} that there 
exists a change of basis $U$ on $\cH$, and integers $d,r$ with $md+r=N$,  
\begin{equation}
U: \;  \cmplxs^N \rightarrow \cmplxs^m \otimes \cmplxs^d\oplus  \cmplxs^r \:, 
\end{equation}
such that, in the resulting representation, all error operators leave 
the factor $\cmplxs^m$ unaffected:
\begin{equation}
U\cA U^\dagger =  \openone_m \otimes \text{Mat}_d(\cmplxs) \oplus \olxs_r \:,
\end{equation}
where $ \olxs_r$ accounts for the action on the remaining summand 
$\cmplxs^r $. Situations where a NS can be directly identified with a 
subset of the physical qubits belong to this category. The above trivial 
two-qubit example, for instance, corresponds to $N=4$, $m=d=2, r=0$. 

\item If $\cA$ is again strictly contained in $\cA_S$, but it cannot be 
identified with the operator algebra of a $m$-dimensional quantum DOF, a 
theorem from the representation theory of operator algebras (the 
Wedderburn theorem, \cite{Arveson,Landsman}; see also \cite{Thirring}) 
still implies that a transformation $U$ to a new basis exists where, as 
above, the action of the errors takes a simple block-factorized form:
\begin{equation}
U: \;  \cmplxs^N \rightarrow \cmplxs^{m_1} \otimes \cmplxs^{d_1} \oplus  
\ldots \oplus   \cmplxs^{m_\ell} \otimes \cmplxs^{d_\ell} \:, 
\label{basis}
\end{equation}
with $\sum_j m_j d_j = N$,  and 
\begin{equation}
U\cA U^\dagger =  \openone_{m_1} \otimes \text{Mat}_{d_1}(\cmplxs) \oplus 
\ldots \oplus  \openone_{m_\ell} \otimes \text{Mat}_{d_\ell} \:.
\label{a}
\end{equation}
Because, in this representation, the action of any error operator in 
$\cA$ is only experienced by the ``syndrome factors'' $\cmplxs^{d_j}$ in the 
decomposition (\ref{basis}), each of the co-factors $\cmplxs^{m_j}$ 
can be identified with the state space of a $m_j$-dimensional NS under 
$\cA$ \cite{Knill-NS, note1}. 
\end{itemize}

It is worth noting that, with respect to the same basis (\ref{basis}), 
the action of operators commuting with the errors becomes ``dual'' to the 
one of the errors themselves,
\begin{equation}
U\cA' U^\dagger =  \text{Mat}_{m_1}(\cmplxs) \otimes \openone_{d_1} \oplus 
\ldots \oplus   \text{Mat}_{m_\ell}(\cmplxs) \otimes \openone_{d_\ell} \:,
\label{aprime}
\end{equation}
implying that a non-trivial transformation is now enforced on the noiseless 
DOFs. Also, whenever $d_j=1$ for some $j$, the syndrome subsystem becomes 
effectively a classical DOF with a one-point configuration space. Thus, 
$ \cmplxs^{m_j} \otimes \cmplxs \simeq \cmplxs^{m_j}$, and the $j$th summand 
in the decomposition (\ref{basis}) can accordingly be identified with a DFS 
under $\cA$ \cite{Knill-NS,Zanardi-Stabilize}.

Because of the abstract algebraic nature of the subsystem identification 
given by (\ref{basis})-(\ref{a}), the mapping between the states of the 
information-carrying NSs and the states of the underlying physical qubits 
may in general become very indirect. 
However, the method for constructing a NS from appropriate observables 
in $\cA'$ may be applied in general \cite{LAScience1}. We now make these 
considerations explicit in the situation that is relevant to the 
experimental implementation.

\subsection{Collective noise for three qubits }

The collective error behavior provides the paradigmatic situation for 
discussing passive noise control through both DFSs and NSs, and in 
particular for realizing the simplest non-trivial noiseless quantum 
subsystem. Collective error models have been extensively analyzed in 
the theory literature 
\cite{Zanardi-NoiselessCodes,Lidar-DFS,Knill-NS,Kempe-FaultTolDFS,Wu-Create},
and experimentally investigated in the QIP context in optical \cite{Kwiat-DFS}, 
trapped-ion \cite{Kielpinski-DFS}, and liquid-state NMR 
\cite{Viola-NS,Fortunato-DFS} 
devices.  For a system $S$ composed of $n$ qubits as above, collective 
noise behavior arises whenever a single environment $E$ couples to the 
individual particles without distinguishing among them.  This results in 
error models that are characterized by {\sl permutation symmetry.}  
Whether or not the natural dynamics of $S$ alone, ruled by $H_S$, actually 
respects this symmetry (as assumed so far) is an important issue with both 
conceptual and practical implications. While deferring a more detailed 
discussion of this point to a later stage, we begin by examining the 
consequences of permutation symmetry.  

In particular, let us focus on the situation where $S$ is composed 
of three qubits, implying that $\cH \simeq \cmplxs^8$,
and the full operator algebra $\cA_S \simeq \text{Mat}_8(\cmplxs)$. 
For the purpose of characterizing collective error models, the crucial 
property is that only {\sl global} error generators 
$J_u= (\sigma_u^{(1)} + \sigma_u^{(2)}+\sigma_u^{(3)})/2$, $u=x,y,z$, may 
be present in the system-environment interaction. By definition, $J_u$ is 
the projection of the total spin angular momentum along the $\hat{u}$-axis 
(in units $\hbar$). 
In NMR with spin-1/2 nuclei, for instance, interactions of this type may 
arise from uniform, fully-correlated magnetic fields which fluctuate in 
direction and strength, leading to evolutions which can be 
semi-classically described as \cite{ErnstBook}
\begin{equation}
|\Psi\rangle_{123} \mapsto |\Psi'\rangle_{123} = e^{-i(\theta_x J_x + 
\theta_y J_y + \theta_z J_z )} |\Psi\rangle_{123} \:, 
\end{equation}
for random variables $\theta_u$, $u=x,y,z$. In the formalism of quantum 
operations, collective error processes are characterized by completely 
positive dynamical maps of the form (\ref{qop}), where the possible 
errors $E_a$ are constrained to commute with all possible particle 
permutations. Thus, the largest interaction algebra $\cA_c$ 
resulting from arbitrary collective interactions consists of all 
the totally symmetric operators on three qubits. Because the dimension of 
the subspace of totally symmetric operators for $n$ qubits is given by 
$(n+1)(n+2)(n+3)/6$ \cite{Zanardi-Sym}, $\cA_c$ is a 20-dimensional 
sub-algebra of $\cA_S$. This implies that a description of the {\sl most 
general} collective error model on three qubits can be accomplished by using 
an error basis with at most 20 (linearly independent) operators, out of the 
possible 64 needed for representing arbitrary noise in the absence of 
symmetries. By writing
\begin{equation}
\cA_c = \text{span} \{ \Sigma_a \, | \, a=0,\ldots, 19  \} \subset 
\cA_S \:, 
\label{Ac}
\end{equation}
an explicit basis of operators $\Sigma_a$ can be constructed by
fully symmetrizing the standard Pauli product operator basis for three
qubits. Let us introduce compact notations to describe operators that 
are invariant under the full set of qubit permutations: 
\begin{eqnarray}
\widehat{ZZ} & = &\sigma_z^{(1)}\sigma_z^{(2)}+\sigma_z^{(2)}\sigma_z^{(3)}
+\sigma_z^{(3)}\sigma_z^{(1)} \:, \nonumber \\
\widehat{ZX}  & = &  \sigma_z^{(1)}\sigma_x^{(2)} + \sigma_x^{(1)}\sigma_z^{(2)} +
   \sigma_z^{(1)}\sigma_x^{(3)} + \sigma_x^{(1)}\sigma_z^{(3)} +
\sigma_z^{(2)}\sigma_x^{(3)} + \sigma_x^{(2)}\sigma_z^{(3)} \:,\nonumber  \\
\widehat{ZZZ} & = &\sigma_z^{(1)}\sigma_z^{(2)}\sigma_z^{(3)} \:, \nonumber \\
\widehat{ZZX} & = & \sigma_z^{(1)}\sigma_z^{(2)}\sigma_x^{(3)} +
\sigma_x^{(1)}\sigma_z^{(2)}\sigma_z^{(3)} +
\sigma_z^{(1)}\sigma_x^{(2)}\sigma_z^{(3)} \:, \nonumber \\
\widehat{XYZ} & = & \sigma_x^{(1)}\sigma_y^{(2)}\sigma_z^{(3)} +
\sigma_y^{(1)}\sigma_x^{(2)}\sigma_z^{(3)} +
\sigma_z^{(1)}\sigma_y^{(2)}\sigma_x^{(3)} +
\sigma_x^{(1)}\sigma_z^{(2)}\sigma_y^{(3)} +
\sigma_y^{(1)}\sigma_z^{(2)}\sigma_x^{(3)} +
\sigma_z^{(1)}\sigma_x^{(2)}\sigma_y^{(3)}\:, 
\label{SymBasis}
\end{eqnarray}
and so forth. Then a basis for $\cA_c$ is given by the $\openone$, the 
three linear operators $J_u$, the six quadratic operators 
$\widehat{XX},\widehat{YY},$ $\widehat{ZZ},\widehat{XZ},\widehat{XY},\widehat{YZ}$, 
and the ten cubic operators resulting from the above construction. 

As noted earlier, the symmetry properties of a given error model appear
explicitly in the commutant of the error algebra. In the case of $\cA_c$,
the commutant $\cA'_c$ contains the subgroup $\Pi$ of unitary operators 
that implement permutations of the particles {\it e.g.}, a swap 
$\pi_{12}$ between qubit 1, 2 means 
$\pi_{12} | i\rangle_1  | j\rangle_2 | k\rangle_3 =  | i\rangle_2  | j\rangle_1 
| k\rangle_3 =| j\rangle_1  | i\rangle_2 | k\rangle_3 $. In fact, one can 
show that the whole $\cA_c'$ consists of linear combinations of operators 
in $\Pi$, expressing  the fact that $\cA_c'$ coincides with the group 
algebra $\cmplxs {\cS}_3$ of the permutation group $\cS_3$ under the above 
representation in $\cH$ \cite{BurrowBook,Zanardi-Stabilize,LAScience1}. 
Physically, because the error generators $J_u$ are also the generators for 
the global rotations of the qubits, $\cA_c'$ can be regarded as containing the 
operators which remain invariant under such rotations. Thus, operators
in $\cA_c'$ can be constructed from the identity and the simplest invariant 
operators \cite{Knill-NS,Viola-Qubit}:
\begin{equation}
s_{12} = \vec{\sigma}^{(1)} \cdot  \vec{\sigma}^{(2)}, \hspace{1cm}
s_{23} = \vec{\sigma}^{(2)} \cdot  \vec{\sigma}^{(3)}, \hspace{1cm}
s_{31} = \vec{\sigma}^{(3)} \cdot  \vec{\sigma}^{(1)} \:,
\label{dot}
\end{equation}
where $\cdot$ denotes the usual dot product. Note that $s_{jk}$ is nothing 
but the Heisenberg spin coupling between spins $j,k$. As it turns out, 
every operator in $\cA_c'$ can be realized, in principle, through the 
application of Heisenberg Hamiltonians of the form (\ref{dot}) 
\cite{Lidar-ThFaultTol}. By observing that the total angular momentum 
observable $J^2=\vec{J}\cdot \vec{J}$ simply rewrites in terms of the above 
operators, collective symmetry immediately implies a conserved quantity, given 
by the eigenvalues $j(j+1)$ of $J^2$.

\subsection{Abelian error models}

Suppose that the system-environment interaction is contributed by a 
{\sl single} global error generator $J_{\hat{v}} = \vec{J}\cdot\hat{v}$, 
$\hat{v}\cdot\hat{v}=1$. Then, in the absence of an independent 
quantizing direction (provided, for instance by $H_S$),
the resulting interaction algebra is abelian, and the corresponding 
error models accounts for collective decoherence (unrecoverable loss of 
phase information) with respect to the fixed basis of eigenstates of 
$\sigma_{\hat{v}}$. The choice $\hat{v}=\hat{z}$ singles out the $z$ 
basis. For three qubits, the relevant error algebra $\cA_z$ can be 
constructed from the generator $J_z$ and the $\openone$, and can be 
identified with the sub-algebra of $\cA_c$ spanned by the four axially 
symmetric and permutation-invariant operators, 
{\it i.e.}
\begin{equation}
\cA_z = \text{span} \{ \openone,J_z,\widehat{ZZ},\widehat{ZZZ} \} 
\subset \cA_c \:,  \label{Az}
\end{equation} 
For instance, an error model in this class that will be of practical
significance is a full-strength (or ``crusher'', borrowing from the NMR 
terminology) collective $z$-dephasing on three qubits, which may be 
described by a quantum operation $\cE_z$ with Kraus operation elements 
$K_a^z$, $a=0,\ldots,3$:
\begin{eqnarray}
K_0^z & =& {1 \over 8} \Big(\openone + 2J_z +\widehat{ZZ} + \widehat{ZZZ} 
\Big) \:, \nonumber \\
K_1^z &= & {1 \over 8} \Big(\openone - 2J_z +\widehat{ZZ} - \widehat{ZZZ} 
\Big) \:, \nonumber \\
K_2^z &=& {1 \over 8} \Big(3\openone + 2J_z -\widehat{ZZ} - 3 \widehat{ZZZ}
\Big) \:, \nonumber \\
K_3^z &= &{1 \over 8} \Big(3\openone - 2J_z - \widehat{ZZ} + 3 \widehat{ZZZ}
\Big) \:.
\label{crusherZ}
\end{eqnarray} 
In a similar way, collective decoherence about the $\hat{x}$ axis can be described 
by effectively switching from the $z$ to the $x$ basis, {\it i.e.} by mapping
$\cA_z \mapsto \cA_x=H\cA_zH$, where $H$ represents a collective Hadamard 
transform. Accordingly, a crusher collective $x$-error process corresponds to 
$\cE_x=\{K_a^x\}$, with operation elements obtained from (\ref{crusherZ}) via
the appropriate rotations.

For later purposes of comparison between encoded and un-encoded information, 
the description of the noise process induced by a three-qubit error model on a 
physical information-carrying qubit will be useful. In general, by 
treating one of the qubits as the data qubit (${\tt d}$) and the remaining 
ones as ancillae (${\tt a}_1$, ${\tt a}_2$), the partial trace operation over 
the ancillae 
\begin{equation}
\rho_{in}= \text{Tr}_{ {\tt a}_1, {\tt a}_2} \{\varrho_{in} \} 
\mapsto \text{Tr}_{ {\tt a}_1, {\tt a}_2 }
\{ \cE ( \varrho_{in}) \} =  \cQ ( \rho_{in} ) 
\label{trace}
\end{equation}
associates to a three-qubit process $\cE$ a one-qubit process 
$\cQ$ on the data qubit ${\tt d}$ alone, provided that the latter is 
initially uncorrelated with the ancillae and the initial state 
$|{\tt a}_1\, {\tt a}_2 \rangle$ is known. For the experimental 
realization, the choice 
$|{\tt a}_1\, {\tt a}_2 \rangle= |0 0\rangle$ will be relevant. 
It is then readily seen that the above $\cE_z$ process corresponds, 
for instance, to applying a one-qubit map of the form
\begin{equation}
\rho \mapsto E^z_+ \rho  E^z_+  + E^z_- \rho  E^z_- \:,
\end{equation}
where  $E^z_\pm = (\openone \pm \sigma_z) /2$ are the usual 
$z$-idempotents, $ (E^z_\pm)^2 =E^z_\pm$. As expected from physical 
intuition, this process is nothing but crusher phase damping on the 
(un-encoded) data qubit.

\subsection{Non-abelian error models} 

Whenever two non-commuting error generators are relevant, the interaction 
algebra describing the resulting error process is non-abelian. In practice, 
we shall be interested at error models obtainable by using 
abelian noise processes as building blocks. 
Let $\cE_x$, $\cE_y$, $\cE_z$ denote crusher dephasing about $\hat{x}$, 
$\hat{y}$, $\hat{z}$, respectively, with associated error algebras $\cA_x$, 
$\cA_y$, $\cA_z$, as above. A simple way for inducing non-abelian error 
processes is through the sequential composition of abelian errors along
different axes. In particular, {\sl crusher isotropic collective 
decoherence} corresponds to cascading crusher noise processes about all 
three axes \cite{Havel-SimDeco}. For instance, $\cE_y \cE_z \cE_x =\cE_{yzx}$, 
with Kraus operators specified by the 64 (linearly dependent) products 
$K_c^y K_b^z K_a^x$, $a,b,c=0,\ldots,3$, of the operators given in the 
previous subsection. While composite noise processes of this sort may not 
naturally occur in physical systems, they can be enforced in liquid-state 
NMR using readily available non-unitary control methods to be described 
later (see also \cite{Havel-SimDeco}).  
In terms of error algebras, the operators describing the composite noise
process $\cE_{yzx}$ can be thought to belong to an error algebra $\cA_{yzx}$ 
which arises from the multiplication of the single-axis algebras, 
\begin{equation}
\cA_{yzx}= \cA_y \cA_z \cA_x = \text{span} 
\{ K_c^y K_b^z K_a^x,\, K_a^x K_b^z K_c^y \} \subseteq \cA_c \:.
\label{composite}
\end{equation}
In fact, one can show that $\cA_{yzx}$ is the full collective algebra 
$\cA_c$ by checking that a basis $\Sigma_a$ of permutation-invariant 
operators (such as the one given in (\ref{SymBasis})) is contained in 
$\cA_{yzx}$. It also turns out that the full $\cA_c$ may be generated 
from the composition of abelian error processes involving two non-commuting 
axes, {\it e.g.} $\cE_z \cE_x =\cE_{zx}$. 

By reasoning as in the abelian case, the one-qubit map $\cQ$ describing 
the effect of a crusher composite noise processes at the single-qubit level
can be derived by evaluating the partial trace (\ref{trace}) on the 
appropriate sets of 16 (for two-axes noise) or 64 (for three-axes noise)
Kraus operators. The result is effective full depolarization on the 
physical data qubit, corresponding to a map of the form
\begin{equation}
\rho \mapsto {1 \over 4} \Big(  \rho  +  \sum_{u=x,y,z}  \sigma_u \rho 
\sigma_u \Big)\:.
\end{equation}

\subsection{ The role of the self-Hamiltonian }

From a physical point of view, it is worth pointing out that while noise 
processes that involve both quantum decoherence (phase damping) and 
dissipation (amplitude damping) always correspond to non-abelian 
error algebras, a non-commutative error algebra does not necessarily 
indicate the presence of genuine energy dissipation in the system. In 
general, in the absence of an internal Hamiltonian, $H_S=0$, every noise
process is effectively equivalent to (adiabatic) decoherence. 
For instance, because all the basic operation elements in (\ref{crusherZ}) 
are Hermitian, crusher collective dephasing as defined above 
is, consistently, a unital process, describing damping of phase 
information in the $z$ basis. Composition of dephasing operations along
non-commuting axes results in depolarization, which is still unital 
and can be regarded as decoherence in all bases.

Because $H_S$ is rarely zero in real systems, the interplay between the 
internal dynamics and the actual error generators $J_a$ turns out to be 
crucial for characterizing the overall open-system dynamics. 
Clearly, it is always possible to construct the interaction algebra 
$\cA$ by including $H_S$, as done so far, among the defining interaction
operators. In particular, the assumption $H_S=\epsilon J_z$ is implicit in 
the original definition of the collective noise model 
\cite{Zanardi-NoiselessCodes}. Regardless of whether $H_S$ commutes or 
not with the $J_a$, if a non-trivial NS is supported by the algebra $\cA$ 
constructed in this way, this NS is completely stable against time evolution.  
Therefore this procedure is, in principle, ideal for devising robust 
quantum memories.
 
However, including $H_S$ among the defining operators for $\cA$ is not 
desirable when the $\cA$ becomes irreducible (thus without useful symmetries) 
or when $H_S$ is regarded as a resource for effecting noise-protected 
manipulations of information. Suppose that an error algebra $\tilde{\cA}$ is 
defined starting from the $\openone$ and the $J_a$ alone. Then $H_S$ may or 
may not belong to $\tilde{\cA}'$. From the point of view of reliable QIP, 
situations falling into the first category are the most favorable, as the 
natural Hamiltonian directly implements a non-trivial logical evolution on 
any NS supported by $\tilde{\cA}$ 
\cite{Fortunato-DFS,Viola-EncodedDecoup,Lidar-EncRecoup}.
If $H_S \not \in \tilde{\cA}'$ instead, then $H_S$ may still preserve a 
{\sl given} NS \cite{Kempe-FaultTolDFS,Fortunato-DFS}, but most likely it 
will have the undesired effect of causing leakage outside the intended 
space. While various schemes are available in principle to cope with 
these effects \cite{NielsenBook,Wu-LEO}, it is not clear to 
what extent these may be viable with realistic control resources. 

In liquid-state NMR systems, collective error models may naturally 
play a role in describing relaxation from fully correlated fields in 
homo-nuclear species \cite{ErnstBook,Redfield}. However, the assumption 
that $H_S$ is proportional to $J_z$ is invalid due to the chemical shift 
effects. For the specific molecule used in the current implementation, 
$H_S$ (to be explicitly given later) will turn out to satisfy none of 
the properties of belonging to the full collective error algebra 
$\cA_c$ or its commutant $\cA_c'$. Nevertheless, the system may be 
used to demonstrate how robustness against $\cA_c$ can be achieved by 
constructing an appropriate NS.

\subsection{The three spin-1/2 noiseless subsystem}

Because, as noted earlier, the eigenvalue $j$ of the total angular momentum 
$J^2$ is conserved, simultaneous eigenstates of $J^2,J_z$ are a natural 
basis to describe the state of the three particles. 
The possible values of $j$ are 
$j=3/2,1/2$, corresponding to a decomposition of $\cH$ as the direct sum 
of two invariant subspaces, $\cH \simeq \cH_{3/2} \oplus \cH_{1/2}$, 
respectively. The quantum numbers $j,j_z$ suffice for completely labeling 
basis states in $\cH_{3/2}$, which is the four-dimensional subspace spanned by 
totally symmetric states ({\it i.e.}, states in $\cH_{3/2}$ transform under 
particle permutations as the one-dimensional symmetric irreducible 
representation of $\cS_3$).
However, with $j=1/2$ and $j_z=\pm 1/2$ this is no longer true for the 
subspace $\cH_{1/2}$, which is also four-dimensional as the eigenvalue 
$j=1/2$ is in fact doubly degenerate. Physically, this degeneracy
accounts for the fact that there are two distinct paths for obtaining a total 
angular momentum $j=1/2$ out of three elementary 1/2 spins. Let $\ell =0,1$ be 
an additional quantum number that labels these two possible paths. Because 
collective errors do not have access to the quantum numbers of the individual 
spins, and the resulting global quantum numbers are the same in both paths, the 
noise can neither distinguish which value of $\ell$ is realized, nor change that 
value. Thus, $\ell$ corresponds to a conserved, two-dimensional DOF under the 
noise. In fact, this is the NS we are seeking \cite{Knill-NS,Viola-Qubit,LAScience2}. 

More formally, basis states in $\cH_{1/2}$ are labeled by two quantum numbers,
$|\ell, j_z\rangle$, with $\ell=0,1, j_z=\pm 1/2$. For fixed $\ell$ 
(fixed path), the resulting subspace carries a copy of the two-dimensional 
irreducible representation of the angular momentum group $su(2)$ 
corresponding to $j=1/2$. There are two such copies, and operators in 
$\cA_c$ do not mix them and act identically on both. 
For fixed $j_z$, one obtains instead a copy of the two-dimensional 
irreducible representation $[2\,1]$ (in Young tableau notation) of the 
permutation group $\cS_3$ \cite{BurrowBook,PeresBook}. 
There are again two such copies, and now operators in $\cA_c'$ do not mix 
them and act identically on both \cite{note3}. 
With respect to the interaction algebra 
$\cA_c$, we can thus identify the DOF corresponding to $\ell$ with a 
subsystem $L$, which is fully protected against errors, and the DOF 
corresponding to $j_z$ with a syndrome spin-1/2 subsystem $Z$, which 
instead experiences the errors. We write $\cH_{1/2} \leftrightarrow \cH_{L} 
\otimes \cH_Z$, and make this identification explicit through the following 
correspondence with states in the computational basis \cite{Viola-Qubit}:
\begin{eqnarray}
\label{EncodedStates}
|{\tt 0}\rangle_L \otimes | +1/2 \rangle_Z &\leftrightarrow &
{1 \over \sqrt{3} }
\Big( |001 \rangle + \omega|010 \rangle + \omega^2|100 \rangle \Big) \:, 
\nonumber \\
|{\tt 0}\rangle_L  \otimes | -1/2 \rangle_Z &\leftrightarrow &
{1 \over \sqrt{3} }
\Big( |110 \rangle + \omega|101 \rangle + \omega^2|011 \rangle \Big) \:, 
\nonumber \\
|{\tt 1}\rangle_L  \otimes | +1/2 \rangle_Z &\leftrightarrow & 
{1 \over \sqrt{3} }
\Big( |001 \rangle + \omega^2|010 \rangle + \omega|100 \rangle \Big) \:, 
\nonumber \\
|{\tt 1}\rangle_L  \otimes | -1/2 \rangle_Z & \leftrightarrow & 
{1 \over \sqrt{3} }
\Big( |110 \rangle + \omega^2|101 \rangle + \omega|011 \rangle \Big) \:, 
\end{eqnarray}
where $\omega = \exp(i2\pi/3)$. Note that two of the totally symmetric 
states spanning $\cH_{3/2}$ are $|000\rangle, |111\rangle$, while the 
remaining two are obtained from (\ref{EncodedStates}) by dropping 
the $\omega,\omega^2$ phase factors. Thus, the full set of basis states 
for $\cH_{3/2}$ and $\cH_{1/2}$ provides an explicit realization of the
general state space decomposition given in (\ref{basis}), with
$U$ being the rotation needed to bring the computational basis to 
the appropriate angular momentum 
basis, and $N=8,m_1=1,d_1=4, m_2=2,d_2=2$, respectively.
Equivalently, the NS qubit living in $\cH_L$ can be identified by 
combining the $\openone$ and the invariant operators (\ref{dot}) into 
three operators $\sigma_u^{(L)} \in \cA'_c$ that behave algebraically
like the Pauli operators. This gives \cite{Viola-Qubit}
\begin{equation}
\sigma_x^{(L)}= {1 \over 2} \Big( \openone +s_{12} \Big) P_{1/2}\:,
\hspace{1cm}
\sigma_y^{(L)}= {\sqrt{3} \over 6} \Big( s_{23} - s_{31} \Big) P_{1/2}\:,
\hspace{1cm}
\sigma_z^{(L)}= i\sigma_x^{(L)} \sigma_y^{(L)} \:, 
\label{obs}
\end{equation} 
where $P_{1/2}= \openone/2 -( s_{12}+ s_{23}+s_{31} )/6$ denotes the 
projector onto the subspace $\cH_{1/2}$. Note that the logical 
$\sigma_x^{(L)}$ observable is simply the restriction to $\cH_{1/2}$ 
of the permutation $\pi_{12}$ swapping qubits 1 and 2.

\subsection{ Collective operators in the noiseless 
subsystem/syndrome representation } 

It is instructive to take a closer look at the action of the error 
generators $J_u$ and of some relevant collective error processes 
directly in terms of the decomposition 
\begin{equation}
U \cH  = \cH_{3/2}  \oplus  \cH_{1/2} \leftrightarrow \cH_{3/2}
\oplus \cH_L \otimes \cH_Z \:,
\label{change}
\end{equation}
where $U$ effects the change of basis mentioned above and 
the subsystem identification within  $\cH_{1/2}$ is explicitly given by 
(\ref{EncodedStates}). By using the identity $1+\omega+\omega^2=0$, 
it is easy to verify that the restriction of the collective noise 
generators $J_u$ to the $\cH_{1/2}$ subspace acts as follows:
\begin{eqnarray}
2J_x P_{1/2} &\leftrightarrow & \openone^{(L)} 
\otimes (-\sigma_x^{(Z)}) \:, \nonumber \\
2J_y P_{1/2} &\leftrightarrow & \openone^{(L)} 
\otimes (-\sigma_y^{(Z)}) \:, \nonumber \\
2J_z P_{1/2} &\leftrightarrow & \openone^{(L)} 
\otimes (+\sigma_z^{(Z)}) \:.
\label{generators}
\end{eqnarray}
Thus, the $J_u$ act as single-qubit errors on the syndrome subsystem 
{\sl alone}. Explicitly, this means that if, for instance, the initial 
state of the system is given by
\begin{equation}
P_{1/2} |\Psi\rangle_{123}=|\Psi \rangle = (\alpha |{\tt 0} \rangle  + 
\beta |{\tt 1}\rangle)_L \otimes 
		(\gamma |+1/2 \rangle + \delta |-1/2\rangle)_Z \:,
\end{equation}
for appropriate coefficients, then the result of a collective rotation 
by, say, $\theta_y$ about $\hat{y}$ is 
\begin{eqnarray}
e^{-i \theta_y J_y} |\Psi\rangle &=& 
(\alpha |{\tt 0}\rangle  + \beta |{\tt 1}\rangle)_L \otimes 
e^{+i \theta_y \sigma_y^{(Z)}/2 } 
(\gamma |+1/2 \rangle + \delta |-1/2\rangle)_Z  \nonumber \\ 
& = & (\alpha |{\tt 0}\rangle  + \beta |{\tt 1}\rangle)_L \otimes 
\Big(  (\gamma \cos(\theta_y/2) +  \delta \sin(\theta_y/2) ) |+1/2 \rangle +
    (\delta \cos(\theta_y/2) - \gamma \sin(\theta_y/2) )  |-1/2 \rangle
\Big)_Z\:,
\label{yrotation}
\end{eqnarray}
while a collective $z$ rotation is simply 
\begin{eqnarray}
e^{-i \theta_z J_z} |\Psi\rangle &=& 
(\alpha |{\tt 0}\rangle  + \beta |{\tt 1}\rangle)_L 
\otimes e^{-i \theta_z \sigma_z^{(Z)}/2 } 
(\gamma |+1/2 \rangle + \delta |-1/2\rangle)_Z  \nonumber \\ 
& = & (\alpha |{\tt 0}\rangle  + \beta |{\tt 1}\rangle)_L \otimes 
 (  \gamma e^{-i \theta_z/2} |+1/2 \rangle +
    \delta e^{+i\theta_z/2}  |-1/2 \rangle)_Z\:,
\label{zrotation}
\end{eqnarray}
and so on. Two observations are worth making. First, if either $\delta$ 
or $\gamma$ is zero, then (\ref{zrotation}) is a direct manifestation of 
the fact that each of the pairs of states in (\ref{EncodedStates}) with fixed 
$j_z$ is a one-qubit DFS against pure $z$ noise \cite{note3}. As explained
in Sect. IIIB, initialization of the ancillae qubits as 
$|{\tt a}_1\,{\tt a}_2\rangle =|00\rangle$ will correpond to encode into
the $j_z=-1/2$ subspace, thus $\gamma=0$ in the implementation. Second, 
Eqs. (\ref{yrotation}) and (\ref{zrotation}) together show that encoding 
into the $Z$ subsystem would instead result in a qubit fully controllable in 
terms of homogeneous local unitaries \cite{Masanes-HSims} {\it i.e.}, 
transformations of the form $U^{(1)}\otimes U^{(2)}\otimes U^{(3)}$ on 
the physical qubits -- for instance, non-selective (``hard'') $\pi$ pulses 
about two non-commuting axes.  

A procedure similar to the one outlined above may be applied to picture 
the effect of an arbitrary, known error model. Take, for instance, crusher 
collective $z$-dephasing with Kraus operators $\cE_z=\{ K_a^z\}$ given in 
(\ref{crusherZ}). One finds that 
\begin{eqnarray}
\widehat{ZZ} P_{1/2} & \leftrightarrow & \openone^{(L)} \otimes 
(-\openone^{(Z)}) \:, \nonumber \\
\widehat{ZZZ} P_{1/2} &   \leftrightarrow & \openone^{(L)} \otimes 
(-\sigma_z^{(Z)}) \:.
\label{zzzzz}
\end{eqnarray}
Thus, when restricted to $\cH_{1/2}$, the action of both 
$K_0^z$ and $K_1^z$ is zero, whereas
\begin{eqnarray}
K_2^z P_{1/2} & \leftrightarrow & \openone^{(L)} 
\otimes E_+^{(Z)} \:, \nonumber \\
K_3^z P_{1/2}  &   \leftrightarrow & \openone^{(L)} 
\otimes E_-^{(Z)} \:.
\label{crusherZ2}
\end{eqnarray}
This just means that the action of crusher collective phase errors 
on the physical system can be pictured as a crusher phase damping 
channel {\sl on the syndrome subsystem $Z$ alone}.
Because, as we shall also comment later, the state of the latter 
abstract subsystem is mapped, upon decoding, into the state of a 
physical ancilla qubit carrying the error syndrome \cite{LAScience1}, 
this means that the physical syndrome subsystem will have experienced 
full phase damping under the same conditions. By the same reasoning, the 
action of crusher composite noise processes can be understood in terms 
of the composition of phase-damping channels affecting the $Z$ subsystem 
along various axes, translating into depolarization of the physical 
syndrome subsystem. These observations will be corroborated by 
experiment.

\subsection{ Verifying infinite-distance error-correcting properties } 

As a result of the above analysis, quantum information encoded in the 
$L$ subsystem is protected indefinitely in time, without requiring any 
active intervention. In the language of QEC \cite{Knill-NS}, this 
stability against the full collective algebra ${\cal A}_c$ characterizes 
$L$ as an infinite-distance QEC code for arbitrary collective errors. 
Formally, this follows from the fact that to the NS one can associate a 
QEC in the usual (subspace) sense by choosing an initial reference state 
$|\varepsilon\rangle_Z$ in the syndrome subsystem (corresponding to 
``no error''), and by letting the code subspace $\cC$ be defined by 
\begin{equation}
\cC =\text{span}\Big\{ |0\rangle_C = |0\rangle_L \otimes 
|\varepsilon\rangle_Z\,, \:
|1\rangle_C= |1\rangle_L \otimes |\varepsilon\rangle_Z \Big \}  \: .
\label{code}
\end{equation}
Then the basis states of $\cC$ verify the necessary and sufficient 
conditions for recovery from all errors in $\cA_c$ \cite{Knill-QEC,Yang-QEC},
\begin{equation}
\langle i_C| E^\dagger_a E_b | j_C \rangle = \alpha_{ab} \delta_{i,j}\:, 
\hspace{1cm} \forall E_a,E_b \in \cA_c \:,
\label{qec}
\end{equation}
for appropriate coefficients $\alpha_{ab}$ -- independent of the logical 
index $i,j$. Note that, {\sl a priori}, infinite-distance behavior as 
expressed by (\ref{qec}) applies regardless of whether $L$ is supported by 
a DFS or by a proper NS. For a DFS, the syndrome state is fixed thus it 
can be effectively disregarded. For a NS, however, the fact that errors are 
allowed to evolve the state $|\varepsilon\rangle_Z$ non-trivially implies 
that the latter may be effectively arbitrary. In both situations, no 
recovery is needed for maintaining information in $L$ \cite{Knill-NS}.

In a NS-QEC experiment, one is interested at inferring robustness properties 
of the encoded information by looking at the code performance under a given set 
of quantum processes. Suppose that, as mentioned in the previous subsection, 
verification is constrained to having initialized the syndrome subsystem in the 
state $|-1/2\rangle_Z$. What kind of conclusions can one draw? The analysis 
is relatively simple under the assumption of perfect implementation 
fidelity. The relevant points can be summarized as follows: 
\begin{itemize}
\item Verifying that quantum information is preserved under the 
implementation of an error process with Kraus operators $\{ K_a\}$ 
implies stability under any error operator $K \in \text{span} \{K_a\}$ 
\cite{Knill-QEC}.  
\item If the set of correctable error operators contains an error algebra 
$\cA$ {\it i.e.}, $\text{span} \{K_a\} \supseteq \cA$, the implementation 
verifies infinite-distance behavior under $\cA$ \cite{Knill-NS}. 
\item Verifying a NS under $\cA$ requires verifying infinite-distance 
QEC for every possible initial state of the syndrome subsystem.
\end{itemize}

These observations can be applied to analyze both abelian and non-abelian
collective error models. For instance, suppose we observe stability under 
crusher $x$ noise. Then we can conclude that operators in $\cA_x$ have identity 
action on $L$ when $Z$ is initialized to $|-1/2\rangle_Z$. However, because the 
state $|+1/2\rangle_Z$ can be reached from $|-1/2\rangle_Z$ by application of 
error operators in $\cA_x$, one has effectively verified a NS against $\cA_x$. 
While a similar argument applies to $y$ noise, the state of the $Z$ subsystem 
is preserved under $z$ noise. Thus, observation of stability under $\cA_z$ for 
fixed initialization in $|-1/2\rangle_Z$ only implies the verification of a 
DFS-behavior under $\cA_z$. A proper NS-behavior under the full collective 
$\cA_c$ can be inferred, in principle, in various ways. 
Keeping the experimental $|-1/2\rangle_Z$ preparation constraint in mind, 
the simplest procedure is to ensure stability under a family of quantum 
processes whose sets of Kraus operators globally span $\cA_c$. 
For instance, one can check that 
\begin{equation}
\text{span}\{ K_a^u K_b^v, K_b^v K_a^u \,|\, a,b=0,\ldots,3 \} = \cA_c 
\end{equation}
for any choice of a composite crusher process which involves two 
non-commuting axes $\hat{u}, \hat{v}$ {\it e.g.}, $\cE_{zy}$ and 
$\cE_{yz}$. The implemented set of processes are discussed in Sect. 
IVB-C. One may notice that the error operators describing $\cE_{yz}$ are 
obtainable as a subset of the errors operators induced by $\cE_{yzx}$. Thus, 
the implemented set is sufficient to infer that a NS for the most general 
collective noise has verified, at least in the limit of sufficiently high 
fidelity. Establishing what minimum fidelity threshold is required for 
inferring NS-verification under non-ideal conditions is a separate 
interesting issue, whose analysis is beyond the scope of this work. 

The three-qubit NS turns out to provide the smallest code capable to 
correct one qubit against the full $\cA_c$. The same, infinite-distance 
protection can be accomplished by using a DFS, but the most efficient DFS 
requires four physical qubits \cite{Zanardi-NoiselessCodes}. 
It is worth noting that, for a given physical system, more efficient 
codes may exist if additional symmetries are present beside the 
permutational one. If, for instance, axial symmetry also applies 
({\it i.e.}, the error model belongs to the class of collective 
single-axis phase damping), then a three-dimensional subspace 
of $\cH$ may be protected with infinite distance by a DFS 
(corresponding to a fixed $j_z$ eigenvalue, $j_z=\pm 1/2$). 
The situation is summarized in Table I.

\section{Experimental design and methods }

\subsection{Liquid-state nuclear magnetic resonance} 

\begin{figure}
\begin{center}
\includegraphics[width=4in]{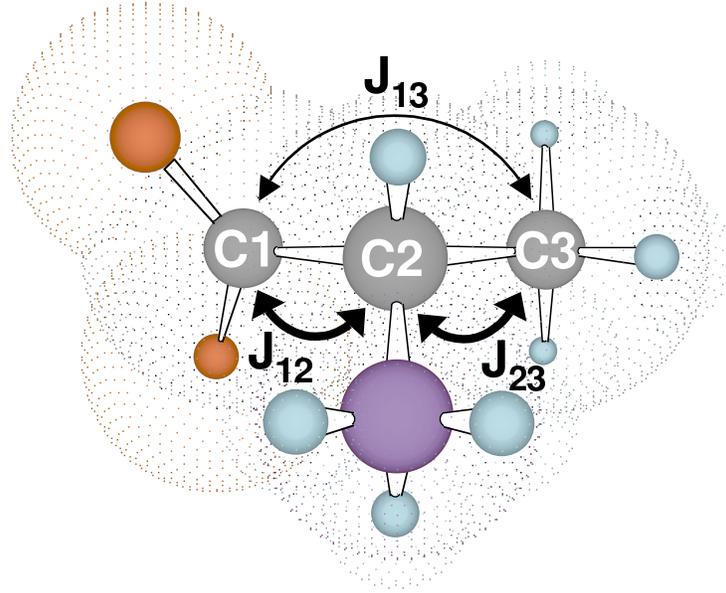}
\caption{ Molecular structure of ${}^{13}$C labeled alanine.
The three carbon spins are used as qubits.  In a reference 
frame that rotates at a frequency 75.4736434 MHz, the internal 
Hamiltonian is accurately described by 
$H_{S}=\pi [\nu_1 \sigma_z^{(1)} + \nu_2 \sigma_z^{(2)}+ 
\nu_3 \sigma_z^{(3)} + (J_{12}\sigma_z^{(1)}\sigma_z^{(2)}+ 
			J_{23}\sigma_z^{(2)}\sigma_z^{(3)} + 
			J_{13}\sigma_z^{(1)}\sigma_z^{(3)})/2)],$
where $\nu_1-\nu_0= 7167$ Hz,  $\nu_2-\nu_0= -2286.5$ Hz, 
$\nu_3-\nu_0= -4881.4$ Hz, $J_{12}=54.1$ Hz, $J_{23}=35.0$ Hz, 
and $J_{13}=-1.3$ Hz.  As it is impractical to directly utilize the
coupling between qubits 1 and 3 due to its weak strength, conditional 
gates involving the pair of qubits $(1,3)$ were effectively replaced 
by sequences of operations involving pairs $(1,2)$ and $(2,3)$ in 
designing and executing quantum networks. The $T_1$ relaxation times 
for the three spins are approximately 21, 2.5, and 1.6 s, while the 
$T_2$ times are about 550, 420, and 800 ms, respectively. 
\label{Alanine} }
\end{center}
\end{figure}

All experiments were performed using liquid-state NMR techniques on a sample 
of ${}^{13}$C labeled alanine (Fig.~\ref{Alanine}) in D$_2$O solution, using a 
300 MHz Bruker {\sc avance} spectrometer.  The spin's evolution is governed by 
the internal Hamiltonian $H_S$, which in the weak coupling limit is 
accurately described by 
\begin{equation}
H_S=\pi \sum_{k=1,2,3}  \nu_k \sigma^k_z + 
	\frac{\pi}{2}\sum_{j>k=1}^3 J_{kj} \sigma_z^{k}\sigma_z^{j},
\label{Hint}
\end{equation}
where $\nu_k$ represents the chemical shift frequency of the $k$th 
spin, and $J_{kj}$ the coupling constant between spins $k$ and $j$, 
respectively (in Hz units). Radio frequency (RF) pulses are used to 
modulate the dynamics to produce the desired net evolution in the spin 
frame where the above internal Hamiltonian is
defined. The interaction with the control field generated from a single 
transmitter has the form \cite{Fortunato-Control}
\begin{equation}
H_{ext}(\omega_{RF},\phi, \omega, t)= \sum_{k} 
e^{-i(\omega_{RF} t+\phi) \sigma_z^k/2 } 
(\omega \sigma^k_x/2) e^{i(\omega_{RF}t+\phi)\sigma^k_z/2 } \:,
\label{Hext}
\end{equation}
where the transmitter's angular frequency $\omega_{RF}$, the initial phase 
$\phi$, and the power $\omega$ are tunable over an appropriate parameter range.  
NMR QIP has been extensively discussed in the literature \cite{Cory-Overview}.  
At room temperature, NMR qubits exist in highly mixed, separable states and so 
NMR QIP relies on ``pseudo-pure'' (p.p.) states whose traceless 
(or ``deviation'') 
component is proportional to that of the corresponding pure state. The identity 
component of the density matrix is unobservable and, if the dynamics are unital, 
constant. The assumption of unital behavior has been validated experimentally,
see Sect. IVD. Under these circumstances, the evolution of a p.p. state 
is equivalent to the one of the corresponding pure state.  
The 3-spin p.p. ground state 
$\varrho_z= |000\rangle \langle000|$ was generated using standard 
gradient-pulse techniques whose details can be found in 
\cite{Teklemariam-EraserPRL}.  
State preparation was experimentally verified by tomographically 
reconstructing the 3-spin density matrix \cite{Chuang-StateTomo}.  A 
constant amount of identity component was added to all reconstructed 
density matrices such that the ground state fidelity with respect to 
the intended 3-spin p.p. state was optimized.

\subsection{Encoding and decoding quantum networks}

\begin{figure}
\begin{center}
\includegraphics[width=\textwidth]{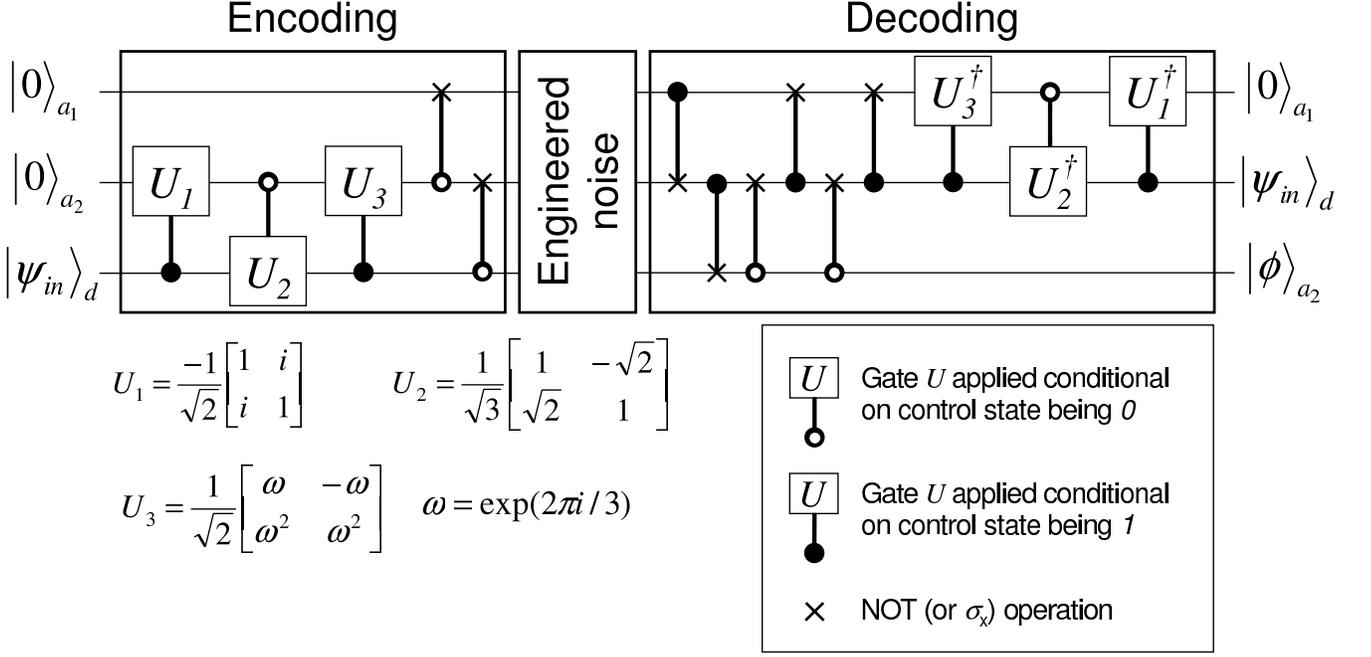}
\caption{ Logical quantum network for the NS experiment.
The information is initially stored in qubit 3, while the ancillae
qubits 1 and 2 are initialized in the state $|0\rangle$.
First, an encoding sequence ($U_{enc}$) is applied in order
to map the initial state of the system into the NS, according to
(\ref{Uenc}).  Next, the qubits are stored in memory while different 
noise processes are applied. Finally, the information is transferred 
to qubit 2 by the decoding sequence ($U_{dec}$).  As noted in the 
text, $U_{enc}$ is a simplified version of $U_{dec}^{-1}$ obtained by 
exploiting the knowledge of the state of the ancillae qubits.
\label{NSNetwork} }
\end{center}
\end{figure}

It is essential that the experimental procedure is designed to allow for 
the protection of an arbitrary one-qubit state, $|\psi_{in}\rangle= 
\alpha |0\rangle + \beta |1\rangle$, with $\alpha$ and $\beta$ potentially 
unknown.  Because collective errors affect the $Z$ subsystem hence 
induce a non-trivial evolution the encoded states, the decoding transformation 
$U_{dec}$ must map the entire set of encoded basis states (\ref{EncodedStates}) 
back to the computational basis properly. Thus, a good decoding transformation
provides an explicit realization of the mapping $U^{-1}$ appearing in 
(\ref{change}). Given $U_{dec}$, an encoding transformation may be 
obtained by letting $U_{enc} = U_{dec}^{-1}$. 
Various choices are possible in principle, differing in the identification they 
establish between the state space of the abstract $L$ and $Z$ qubits and the 
physical qubits. The choice of $U_{dec}$ we implemented maps $L$, $Z$ into the 
data qubit 2 and the ancilla qubit 3, respectively:
\begin{eqnarray}
 |{\tt 0}\rangle_L \otimes | +1/2 \rangle_Z  
\leftrightarrow  |j=1/2, \ell=0, j_z=+1/2 \rangle
&\mapsto &|001\rangle \nonumber \:\\
 |{\tt 1}\rangle_L \otimes | +1/2 \rangle_Z 
\leftrightarrow  |j=1/2, \ell=1, j_z=+1/2 \rangle
&\mapsto &|011\rangle \nonumber \:\\
 |{\tt 0}\rangle_L \otimes | -1/2 \rangle_Z 
\leftrightarrow  |j=1/2, \ell=0, j_z=-1/2 \rangle
&\mapsto &|000\rangle \nonumber \:\\
 |{\tt 1}\rangle_L \otimes | -1/2 \rangle_Z 
\leftrightarrow  |j=1/2, \ell=1, j_z=-1/2 \rangle
&\mapsto &|010\rangle \:.
\label{Udec}
\end{eqnarray}
The fact that (\ref{Udec}) only specifies $U_{dec}$ on the $\cH_{1/2}$ 
subspace is reflected by the fact that the value of the first ancilla qubit 
remains set to zero. $U_{dec}$ is uniquely determined by also defining its 
action on the $|j=3/2,j_z \rangle$ states spanning $\cH_{3/2}$ or, 
equivalently, the mapping with the remaining four computational basis 
states:
\begin{eqnarray}
 |j=3/2,j_z=+3/2 \rangle &\mapsto &\;\;\:\,\,|100\rangle \nonumber \:\\
|j=3/2,j_z=+1/2 \rangle &\mapsto  &-i|111\rangle \nonumber \:\\
 |j=3/2,j_z=-1/2 \rangle &\mapsto &-i|110 \rangle \nonumber \:\\
 |j=3/2,j_z=-3/2 \rangle &\mapsto &\;\;\:\,\,|101 \rangle \:.
\label{Udec2}
\end{eqnarray}
Having determined $U_{dec}$, no general efficient procedure is known 
for designing a logical network of realizable one- and two-qubit operations 
effecting $U_{dec}$. 
The implemented networks for $U_{dec}$ and $U_{enc}$ are shown in 
Fig.~\ref{NSNetwork}.  Due to the weak strength of the $J_{13}$ coupling, no
direct gates between spins 1 and 3 are used.  For practical convenience, 
$U_{enc}$ is simplified by using qubit 3 as the information carrier 
({\it i.e.}, switching qubits 2 and 3 with respect to the output) and by 
taking advantage of the knowledge of the starting state $|00\rangle$ of the 
input ancillae.  Thus, the implemented $U_{enc}$ effects the transformation
\begin{equation}
U_{enc} |00\psi_{in} \rangle \leftrightarrow(\alpha |{\tt 0} \rangle_L 
\otimes 
|-1/2\rangle_Z + \beta |{\tt 1} \rangle_L \otimes |-1/2 \rangle_Z) =
|\psi_{in} \rangle_L \otimes | -1/2 \rangle_Z \:.  
\label{Uenc}
\end{equation} 

The logical gates involved in the NS encoding and decoding circuits were 
first mapped into ideal pulse sequences via standard quantum network 
methods~\cite{Cory-Overview}.  Pulses were then implemented by strongly 
modulating the internal Hamiltonian (\ref{Hint}) of the alanine molecule 
with externally controlled RF magnetic fields as mentioned above.  
Details of pulse design can be found in \cite{Fortunato-Control}.  
Combinations of rotations that were used in multiple places in the 
sequence were merged into a single pulse and directly implemented.  
In practice, inaccurate preparation of the ancilla state $|{\tt a}_1 
{\tt a}_2 \rangle =|0 0\rangle$ and operational errors may result in 
unintentionally populating states in the $\cH_{3/2}$ subspace. 
Given the decoding action (\ref{Udec2}), this could contribute to 
observable signal upon discarding qubits 1 and 3. The fact that contributions 
originating from $\cH_{3/2}$ remained negligibly small was inferred in the 
implementation from the absence of appreciable double- and triple-coherence 
decay modes and from the stability of the observed signal against applied 
noise strength \cite{Viola-NS} (See also Sect. IVB). 

It is worth emphasizing the difference between the ability to encode an 
{\sl arbitrary} quantum state and the ability to accomplish a desired state 
preparation. While the latter is appropriate in the context of initializing 
a quantum algorithm, the former is crucial for quantum memory purposes 
(see also \cite{Kielpinski-DFS}) and for QEC in general. In the specific 
case, much simpler procedures would suffice for initializing the system 
into a desired, {\sl known} NS state. For instance, one can observe that 
with an appropriate choice of basis a NS always contains a state which 
is the tensor product between a two-qubit singlet and a one-qubit 
logical state (see \cite{Viola-Qubit} for explicit encoded states 
alternative to (\ref{EncodedStates})). Then, similar to the case of 
initialization in a DFS \cite{Zanardi-SemiCond1}, synthesizing a NS 
state may be achieved by relying on a unitary transformation that 
prepares singlet states. Alternatively, one could exploit appropriate 
non-unitary control such as cooling, as suggested in 
\cite{DiVincenzo-Exchange}. For either DFSs or NSs, the existence 
of such initialization procedures does not automatically translate into 
the existence of an efficient way for effecting a general encoding. 
From this point of view, a systematic comparison of network complexity 
for DFSs vs NSs is worth being examined in more detail, and will be 
addressed elsewhere.

\subsection{ Heisenberg representation of collective noise }

As discussed in Sect. IIA, it is unlikely that noise symmetries 
directly imply the preservation of one (or more) of the natural 
subsystem's DOF (such as the states of the physical spins).  Yet, 
an abstract information-carrying subsystem not identifiable with 
any of the physical qubits exists through the subsystem identification 
(\ref{EncodedStates}) \cite{Viola-Qubit, LAScience1}. 
The decoding operation corresponds to extracting the abstract protected 
DOF by mapping it to the natural DOF associated with the data qubit.  
In the abstract NS/syndrome representation, the desired error-correcting 
behavior translates into the property that collective errors act only
on the $Z$ subsystem, as discussed earlier and explicitly verified 
in (\ref{generators}).
An equivalent description can be constructed directly in terms of the 
physical DOFs by looking at the error algebra $\cA_c$ in an appropriate
Heisenberg representation determined by $U_{dec}$.  

Let $E_a \in \cA_c$ be any collective error operator. 
The state of the entire system after it has been encoded into the NS, 
affected by the collective error, and then decoded from the NS is given by
\begin{equation}
|\psi_{out} \rangle_{\tt d}|0\rangle_{{\tt a}_1}|\phi_{out}
\rangle_{{\tt a}_2} = 
 U_{dec} E_a 
U_{dec}^\dagger |\psi_{in}\rangle_{\tt d}|0\rangle_{{\tt a}_1}
|\phi_{in}\rangle_{{\tt a}_2} = E_a^H 
|\psi_{in}\rangle_{\tt d}|0\rangle_{{\tt a}_1}|\phi_{in}
\rangle_{{\tt a}_2} \:,
\label{heisenberg}
\end{equation}
where the Heisenberg-transformed error operator $E_a^H$ is defined by
$E_{a}^H= U_{dec} E_a U_{dec}^\dagger$. Because  $E_a$ is in $\cA_c$, 
the action of $E_a^H$ can be inferred by knowing the transformed 
collective generators $J_u^H$, $u=x,y,z$.  
Using the decoding network of Fig.~\ref{NSNetwork}, and recalling that 
${\tt d}$, ${\tt a}_1$, ${\tt a}_2$ denote the data, the first and second
ancilla qubit, respectively, one finds
\begin{eqnarray}
2 J_x^H &=&  E_+^{{\tt a}_1} (- \sigma_x^{{\tt a}_2}) \openone^{\tt d} + 
	E_-^{{\tt a}_1} \sigma_x^{{\tt a}_2} 
	(\openone^{\tt d}-2\cos(\pi/3)\sigma_y^{\tt d}-2\sin(\pi/3)
        \sigma_z^{\tt d}) \nonumber \\
2 J_y^H &=&  E_+^{{\tt a}_1} (- \sigma_y^{{\tt a}_2}) \openone^{\tt d} + 
	E_-^{{\tt a}_1} \sigma_y^{{\tt a}_2} 
	(\openone^{\tt d}-2\cos(\pi/3)\sigma_y^{\tt d}-2\sin(\pi/3)
        \sigma_z^{\tt d}) \nonumber\\
2 J_z^H &=& E_+^{{\tt a}_1} (+ \sigma_z^{{\tt a}_2}) \openone^{\tt d} + 
	E_-^{{\tt a}_1} \sigma_z^{{\tt a}_2} (\openone^{\tt d}
        +2\sigma_z^{\tt d}) \:.
\label{JHRep}
\end{eqnarray}
The above equations make it explicit that the action of any collective 
error is identity on the data bit, provided that the state of the first 
ancilla ${\tt a}_1$ qubit is properly set to $|0\rangle$.  This identity 
action is the counterpart of the one expressed in the Schr\"{o}dinger 
picture by Eqs. (\ref{generators}). On the other hand, the second ancilla 
qubit ${\tt a}_2$ is in general evolved non-trivially by the noise. According 
to (\ref{JHRep}), the protection afforded by the NS can be equivatently 
understood as a rotation of the error algebra such that the abstract 
protected DOF is identified with the data qubit {\sl conditionally} to the 
state of the first ancilla qubit.

\subsection{ Design of the evolution period }

In order to verify the behavior of the NS code in a controlled setting, 
the delay period between encoding and decoding needs to be carefully 
designed so as to implement an effective evolution under a precisely 
known error model. An important building block for the procedure is the 
ability to implement a ``no-op'' evolution {\it i.e.}, an evolution 
corresponding to the identity operation that provides the reference of 
no applied errors.

\begin{figure}
\begin{center}
\includegraphics[width=3in]{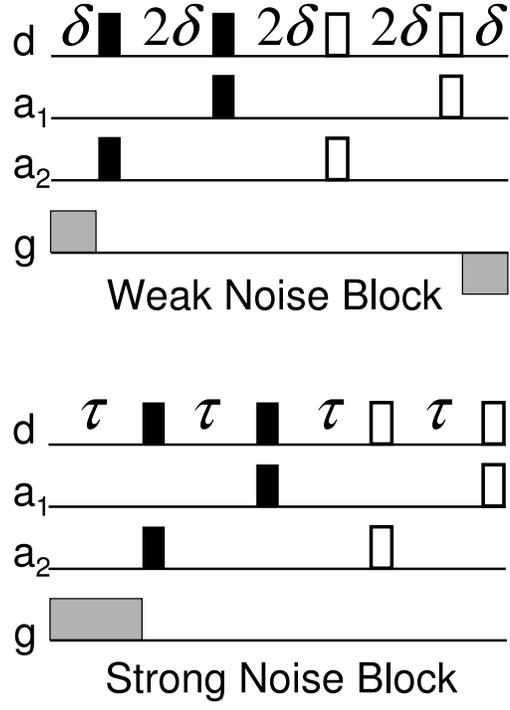}
\caption{No-op pulse sequences for both weak and strong noise applied
to the data qubit ({\tt d}) and the two ancillae qubits ({\tt a}$_1$ 
and {\tt a}$_2$).  Black (white) boxes represent $\pi$ pulses about 
the $\hat{x}$ ($-\hat{x}$) axis.  
The experimental delay times were $\delta \sim 5.5$ ms and 
$\tau \sim 0.5$ ms.  The bottom line ({\tt g}) denotes the magnetic 
field gradient sequence used to introduce either weak (top) or strong 
(bottom) noise.  As explained in Sect. IIID, all gradients were along 
the $\hat{z}$ axis and were only applied during the free evolution 
sub-interval of the whole evolution period.  Collective noise along 
other axis was implemented by collective rotations sandwiching the entire 
noise block.  Cascades of noise along non-commuting axes were implemented
by applying multiple noise blocks in series.
\label{NSNoiseBlocks} }
\end{center}
\end{figure}

Additional constraints exist because, as mentioned, the assumptions 
made on $H_S$ in Sec.~\ref{Theory} are not achievable in the current 
experimental setting.  In particular, $H_S$ does not itself respect 
the permutational symmetry of the collective error model, neither 
does it belong to the commutant $\tilde{\cA}_c'$ of the collective 
error algebra, causing the system to depart from the protected NS.  
The net, unwanted evolution induced by $H_S$ over the delay period
can be averaged out by using refocusing techniques 
\cite{ErnstBook,jones}. The implemented sequences of $\pi$ pulses are 
depicted in Fig. \ref{NSNoiseBlocks}. 

Because the system evolves through intermediate states outside the NS 
during the refocusing cycle, protection of the quantum data cannot be
expected if noise is acting over the entire evolution period. However, 
what is crucial for verifying the robustness of the intended NS is that 
the information resides {\sl inside} the protected space while the noise 
is applied. This can be achieved by making sure that the basic building 
block for engineering collective noise processes is a $z$-gradient, 
in which case $H_S \in \cA'_z$, and by applying the noise only during
the sub-interval of the decoupling cycle corresponding to the identity 
frame ({\it i.e.}, to the free evolution sub-interval) 
\cite{ErnstBook,Viola-Control,note2}. While this suffices for testing 
collective dephasing processes along $\hat{z}$, error algebrs of the 
form $\cA_x$, $\cA_y$ can also be probed, by simply sandwiching a 
$z$-noise process with the appropriate collective rotation pulse about 
either the $\hat{x}$ or $\hat{y}$ axis \cite{Havel-SimDeco}. Finally, 
cascades of noise blocks involving different axes were implemented by 
applying multiple noise blocks in series during a longer evolution period.

\subsection{Implementation of collective error models}

Two different engineered noise models were implemented to test both the 
weak and strong noise limits \cite{Viola-NS}.  Both relied on linear 
magnetic field gradients in order to create an incoherent evolution over 
the spatial distribution of the sample \cite{Sodickson-KSpace}.  
The net phase evolution caused by a gradient pulse for a spin located 
$\delta_z$ from the center of the sample is given by
\begin{equation}
\Delta \Phi (\delta_z) = \exp\Big(-i \gamma \frac{dB_z}{dz} 
\delta_z \Big)\:,
\end{equation}
where $\gamma$ is the gyro-magnetic ratio of the nuclear species and 
determines how strongly the magnetic moment of the spin couples to the 
magnetic field.  A strong gradient pulse causes spins at the edge of the 
sample to evolve through many cycles, producing an almost uniform
distribution of phases across the sample.  
In the absence of molecular motion, the effects of this incoherent 
evolution could be refocused by a reverse gradient pulse. However, in 
liquid samples at room temperature, the spins are undergoing a random
spatial diffusion that renders this incoherent process effectively 
irreversible, resulting in unrecoverable loss ({\it i.e.} decoherence) 
of the quantum information.  The exponential attenuation of the signal
associated with the combined gradient-diffusion process is described 
by a factor~\cite{Sodickson-KSpace},
\begin{equation}
A=\exp\Big(-D \int k^2(t) dt\Big)\:,
\end{equation}
where $D$ is the molecular diffusion coefficient and 
$k(t)=\gamma t (dB_z/dz)$.  For a gradient pulse of duration $\delta$, 
and a diffusion period of duration $\Delta$ followed by an inverse 
gradient also of duration $\delta$, the total attenuation of the 
coherence reduces to
\begin{equation}
A=\exp\Big( -D \gamma^2 ({dB_z}/{dz})^2 \delta^2 
(\Delta+\frac{2 \delta}{3})\Big)\:.
\label{Grad-Diff}
\end{equation}
Therefore, one can associate this evolution with a $T_2$-like process,
$ \exp(-t/{\tau})$, where $1/\tau$ represents the relevant, effective 
noise strength.  By equating $ A = \exp(-t/\tau)$, the elapsed 
time $t= \Delta+2 \delta$ being the duration of the gradient-diffusion 
process, the noise strength is given by
\begin{equation}
\frac{1}{\tau} = D \gamma^2 (dB_z/dz)^2 \delta^2 \,
        \frac{ \Delta+\frac{2 \delta}{3} }{\Delta+2 \delta}\:.
\label{tau}
\end{equation}
This reduces to the simpler expression quoted in \cite{Cory-3BitQEC} if
$2\delta \ll \Delta.$ According to (\ref{tau}), the noise strength can be 
tuned by changing either the holding duration $\Delta$ or the gradient 
strength $dB_z/dz$.  In order to keep the effects of other natural 
decoherence mechanisms fixed in different experimental runs, only the 
gradient strength was varied in practice ({\sl decoherent implementation}).  
Each effective noise rate $1/\tau$ was independently measured using a 
stimulated echo sequence~\cite{Sodickson-KSpace}.  

If no reverse gradient is applied to refocus the magnetization, then the 
evolution remains incoherent until the ensemble signal is acquired, at 
which time the spatial degrees of freedom are traced over rendering the 
evolution effectively irreversible.  Therefore, the crusher noise 
limit can be probed by using a strong gradient pulse with no reverse 
gradient pulse ({\sl incoherent implementation}). 
The details of the no-op sequences and the different noise 
implementations are given in Fig.~\ref{NSNoiseBlocks}.

\section{Results}

\subsection{ Metric of control }

We used the entanglement fidelity~\cite{Schumacher-Channels} as the 
reliability measure quantifying how well quantum information 
was preserved under the evolution of an implemented super-operator 
$\cal{Q}$.  Given an operator-sum decomposition of $\cal{Q}$ in terms 
of Kraus operators $\{ A_\mu \}$, and an input state $\rho$, the 
entanglement fidelity can be computed as
\begin{equation}
F_e (\cQ, \rho) = \sum_\mu Tr(\rho A_\mu) Tr(\rho A_\mu^\dagger) \:.
\label{SchumacherFidEnt}
\end{equation}
For a uniform distribution of input states, the maximally mixed input 
state, $\rho = \openone/2$, is used to characterize the un-biased channel 
performance.  The entanglement fidelity can then be inferred from 
experimentally available data once it is expressed in terms of either 
pure state fidelities or relative input-output spin polarizations.
Under the additional assumption of unital dynamics, {\it i.e.,} 
$\cal{Q}(\openone) = \openone$, the relevant expressions for 
single-qubit quantum process tomography are, respectively:
\begin{equation}
F_e (\cQ) = {1 \over 2} 
\Big(F_{|+x\rangle} + F_{|+y\rangle} +F_{|+z\rangle} - 1\Big)\:,   
\label{InOutFidEnt}
\end{equation}
or, equivalently, 
\begin{equation}
F_e (\cQ) = {1 \over 4} \Big(1+ p_x + p_y +p_z \Big)\:. 
\label{AvgCorr}
\end{equation}
In (\ref{InOutFidEnt}), the pure state input-output fidelity 
$F_{|\psi\rangle} = \mbox{Tr}\{ |\psi\rangle \langle\psi| 
\cal{Q}(|\psi\rangle \langle\psi|) \}$, and $|u \rangle$,  
$u=x,y,z$, are the eigenvectors of the corresponding Pauli operator 
with positive eigenvalue (thus, 
$|+z \rangle = |0\rangle$, $|-z \rangle = |1\rangle$)
\cite{Knill-5BitQEC,Viola-NS}.  In (\ref{AvgCorr}), 
$p_u = \mbox{Tr}\{ \sigma_u \cQ(\sigma_u) \}/2$
represents the relative output polarization given input $\sigma_u$ 
\cite{Cory-3BitQEC,Knill-5BitQEC,Fortunato-DFS}. The implementation
allowed to explicitly validate the consistency of the above expressions
within experimental accuracy.

\subsection{ Weak collective noise along a single axis }

\begin{figure}
\begin{center}
\includegraphics[width=5in]{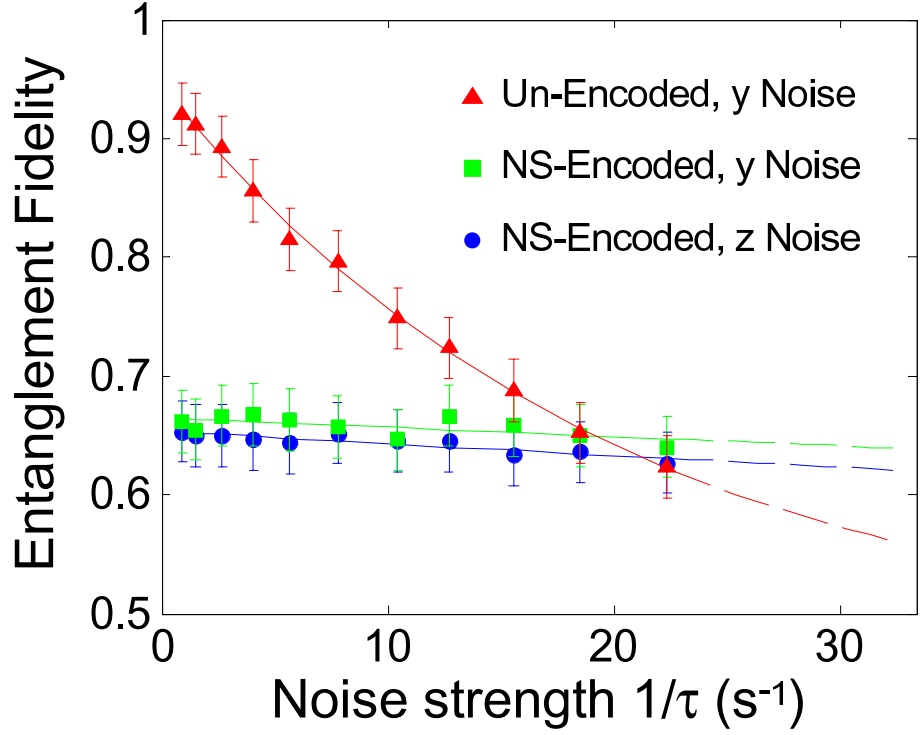}
\caption{Measured entanglement fidelities for single-axis 
collective noise in the weak noise regime along either the 
$\hat{y}$ axis (NS-encoded (squares) and un-encoded data (triangles)), 
or the $\hat{z}$ axis (NS-encoded data only (circles)).  The decay 
of the un-encoded spin, $C_3$, was obtained by turning off the 
encoding and decoding sequences and by subjecting it to the noise 
sequence alone.  The data are fit to an exponential decay function, 
with the interpolated (solid) and extrapolated 
(dashed) lines shown in the plot.  
\label{NSWeakNoiseData}}
\end{center}
\end{figure}

The ability of the NS to protect quantum information against weak 
collective noise was tested using the gradient-diffusion techniques 
described above.  The entanglement fidelity of the data qubit was 
experimentally determined from Eq. (\ref{InOutFidEnt}) as a 
function of noise strength for single axis collective noise. Separate
experiments were carried over for extracting the behavior of both 
the NS-encoded qubit and the un-encoded data spin. The results are 
summarized in Fig.~\ref{NSWeakNoiseData}.  The measured curves are 
fitted to a decaying exponential function, 
$F_e = A \exp(-t/\tau) +B$, as expected for a dephasing channel 
induced by a single-axis noise mechanism.  The un-encoded 
data's fit is characterized by $A=0.51 \pm 0.04$ and $B=0.43 \pm 0.03$, 
confirming the expected decaying contribution.  The NS data's fit is
instead characterized by $A=0.03 \pm 0.03$ for both $y$ and $z$ noise,
with constant coefficients $B=0.62 \pm 0.02$ and $B=0.64 \pm 0.02$, 
respectively.  

For all implementations, departures from the expected ideal behavior
may be explained by pulse imperfections as well as by naturally
occurring relaxation processes, whose effects are assumed to be 
independent of the applied noise strength.  
First, the fact that the decaying contribution (coefficient $A$) is 
small for both of the encoded situations indicates that the data resides
in the NS during the application of the noise, which makes it insensitive 
to collective errors. 
Second, the fact that the constant term (coefficient $B$) is greater 
than $0.50$ \cite{Bennett-MixedQEC}, confirms that quantum information is 
retained, in principle, for arbitrary noise strengths despite significant 
imperfections in the implementation of the encoding and decoding operations.  
Finally, the extrapolation of the data bounds the 
performance of the NS in the strong noise limit. 

\begin{figure}
\begin{center}
\includegraphics[width=5in]{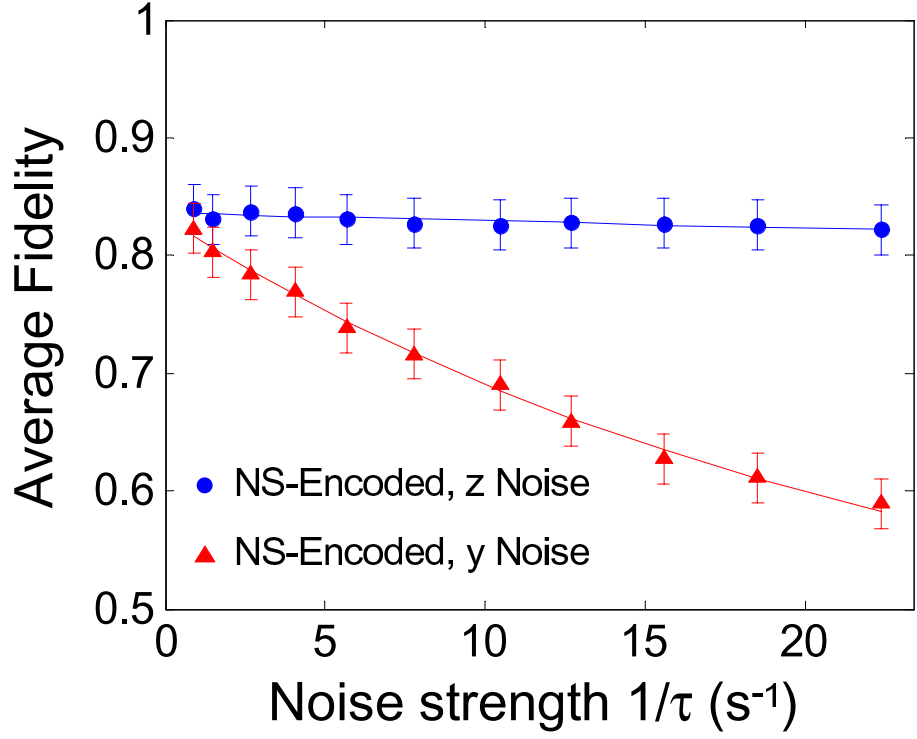}
\caption{Measured average state fidelity for the second ancilla 
qubit, ${\tt a}_2$, for single-axis collective noise in the weak regime.  
Because the encoding transformation $U_{enc}$ maps the initial state 
into an eigenstate of $J_z$ (see (\ref{Uenc})), the ancilla qubit's 
state is unchanged by the application of noise processes along 
$\hat{z}$. This demonstrates that the encoding also maps the system 
into a one-qubit DFS against collective $z$-dephasing \cite{note3}.
The case of collective $y$ noise demonstrates, however, that 
information is not stored in an eigenstate of a generic collective 
noise operator.
\label{NSWeakNoiseExtraBitData} }
\end{center}
\end{figure}

As discussed in Sect. IIG-H, unlike the case of a noiseless subspace the 
encoded NS states are not eigenstates of all noise operators.  This 
implies, in general, a decay of the full-state fidelity due to evolution 
of the ancillae qubits under the action of the noise.  
Because the system always resides, within the experimental accuracy, in 
the spin-1/2 subspace, the ancilla ${\tt a}_1$  remains unchanged by the 
application of the noise while the ancilla qubit ${\tt a}_2$, which is 
mapped from the syndrome $Z$, is decohered by the action of $y$ noise 
operators. For the case of collective $z$ noise, and with initialization 
in a fixed $j_z$ subspace, the system resides into a DFS, as noticed 
repeatedly. Thus, the state of the second ancilla qubit remains also 
unchanged under the $z$-dephasing process. These behaviors are experimentally 
confirmed in Fig.~\ref{NSWeakNoiseExtraBitData}. Note that because $|{\tt a}_2
\rangle$ is always initialized to $|0\rangle$, $F_e$ does not provide an 
appropriate metric. An average input-output fidelity is evaluated instead, 
resulting from a uniform average over the data input states.

\subsection{ Incoherent implementation to mimic strong collective noise }

A variety of incoherent collective noise processes were also implemented 
to explore the strong noise limit and to establish robustness of the 
implemented NS against the full, non-abelian $\cA_c$.  The entanglement 
fidelity data, again calculated via Eq. (\ref{InOutFidEnt}), are presented 
in Table \ref{StrongNoiseData} \cite{Viola-NS}.
As expected, the un-encoded qubit's entanglement fidelity drops to $0.50$
for single axis noise (corresponding to a full-strength dephasing channel),
and to $0.25$ for a cascade of non-commuting noise blocks (inducing a 
full-strength depolarizing channel).  For the NS-encoded case, 
the entanglement fidelity again departs from unity due to pulse 
errors and non-collective natural decoherence mechanisms, but it 
is changed only slightly under the action of different noise blocks.
The significant improvement in the amount of information retained under 
a cascade of strong noise mechanisms confirms the expected benefits of the 
NS encoding already indicated by the weak noise data.

\subsection{ Experimental determination of channel super-operators }

To gain further insight about the quantum processes realized in the 
experiment, the direct reconstruction of the super-operators describing 
some representative crusher dephasing channels was also obtained, leading 
to explicit sets of single-qubit Kraus operators.
The measurement of the input-output relations for the set of input 
states $|+x\rangle\langle +x|$, $|+y\rangle\langle +y|$, 
$|+z\rangle\langle +z|$, and $|-z\rangle\langle -z|$ is sufficient to 
allow the determination of a Kraus form for a desired super-operator
\cite{Childs-ProcTomo,Havel-SuperOp}.  While this analysis assumes that 
the channel is trace-preserving, unitality is now not assumed, thereby
allowing to independently test this assumption.

Table~\ref{SuperOperator} collects the experimentally determined 
one-qubit super-operators and corresponding entanglement fidelities,
calculated via Eq. (\ref{SchumacherFidEnt}) ($\rho = \openone/2$) 
for five representative channels. Because the diagonal elements in 
each of the channel super-operators measure the relative polarizations
$p_u$ introduced in (\ref{AvgCorr}), $F_e(\cQ)$ can equivalently 
be evaluated from the trace of the relevant super-operator 
representation (up to a factor 1/4).

First, the entanglement fidelity values are consistent with the ones 
calculated via Eq. (\ref{InOutFidEnt}).  Second, the data show that the 
unitality assumption is broken by an average deviation no larger than 
$5\%$.  Finally, it is worth stressing that the measured super-operators 
contains complete information on the relative contributions of coherent 
vs decoherent errors.  For instance, the purity of an output state 
$\rho_{out}$, given by $\xi = \mbox{Tr}\{ \rho_{out}^2\}$ 
for a given input state ${\openone, \sigma_x, \sigma_y, \sigma_z}$ 
can be inferred from the sum of the squares of the elements of each 
row of the super-operator.

\section{Conclusions}

We presented a thorough theoretical and experimental investigation 
of the significance of a NS in the context of quantum information 
protection. This exploration demonstrates both the utility of the NS 
and its limitations. In particular, the assumption that the internal 
Hamiltonian is either proportional to one of the noise operators or
that it is trivially zero is not valid for the NMR implementation, 
and is unlikely to be met exactly in any experimental realization of 
a quantum information processor.  If such assumptions are at least 
approximately met, NSs represent the most efficient means for 
protecting quantum information in the presence of noise interactions 
with dominant symmetry components.  For the example of the 3-qubit 
collective noise model, our experimental results convincingly 
demonstrate improvements in protecting information against a class 
of both abelian and non-abelian collective error models.  
Finally, experimentally reconstructed single-qubit super-operators 
provide detailed information about the properties and the relative 
contributions of coherent and decoherent errors in the implementation.

\section{Acknowledgments}

This work was supported by the National Security Agency and Advanced
Research and Development Activity under Army Research Office contract number
DAAD19-01-1-0519, by the Defense Sciences Office of the Defense Advanced
Research Projects Agency under contract number MDA972-01-1-0003, and by
the Department of Energy under contract W-7405-ENG-36. L.V. also 
gratefully acknowledges support from a J. R. Oppenheimer Fellowship. 
We thank Grum Teklemariam and Nicolas Boulant for help with implementation, 
and Greg Boutis, Roberto Onofrio, and Seth Lloyd for valuable discussions.

\bibliography{NSLong}

\begin{thebibliography}{66}
\bibitem[{not({\natexlab{a}})}]{note1}
\bibinfo{note}{The ``identity action'' of the relevant error operators on a DOF
  of the system, as expressed by (\ref{a}), remains the key (necessary and
  sufficient) condition for characterizing NSs without reference to a
  $\dagger$-closed algebraic structure as invoked here. Such generalization may
  be relevant to handle open-system evolutions directly described in terms of
  master equations, see for instance \cite{Kempe-FaultTolDFS}.}
\bibitem[{not({\natexlab{b}})}]{note3}
\bibinfo{note}{For fixed $j_z$, each of the two-dimensional subspaces spanned
  by $\{ |\ell, j_z\rangle, \,\ell=0,1 \}$ is a one-qubit DFS under the
  collective dephasing error algebra $\cA_z$.}
\bibitem[{not({\natexlab{c}})}]{note2}
\bibinfo{note}{Note that, if desired, application of the noise during the
  entire evolution period could be effected at the expenses of modulating the
  gradient Hamiltonian during each decoupling sub-interval, so as to compensate
  for the rotations due to the refocusing pulses. See \cite{Viola-Control} for
  related ideas on fast-switched control schemes.}
\end{thebibliography}

\newpage

\begin{table}
\begin{center}
\begin{tabular}{|cccccc|}
\hline
Type of errors & Strength & Symmetries & Dim$_{\cA}$ & 
Error control code & Dim$_{\cC}$  \rule[-2mm]{0mm}{7mm}\\ \hline \hline
General independent & Arbitrary & None & 64	& None & None \\
General independent & Weak      & None & 10	& None & None \\
Axial independent   & Arbitrary & Axial & 8	& None & None \\
Axial independent   & Weak      & Axial & 4	& 3bit QEC & 2 \\
\hline
General collective  & Arbitrary & Perm & 20	& 3bit NS  & 2 \\
General collective  & Weak      & Perm & 4	& 3bit NS  & 2 \\
Axial collective    & Arbitrary & Perm, axial & 4 & 3bit DFS & 3 \\
Axial collective    & Weak      & Perm, axial & 2 & 3bit DFS & 3 \\ 
\hline \hline 
\end{tabular}
\caption{Comparison between relevant error models on 3 qubits and 
corresponding error control strategies. The various columns list, 
in the order: the type of environment interaction (independent or 
collective); the relevant noise strength (or, equivalently, the 
order in time at which protection is sought \cite{Knill-NS}, weak 
strength corresponding to first-order effects); the type of symmetry
(permutation or axial); the dimension of an error basis for the 
appropriate error set; the most efficient error control code 
available for implementation on 3 qubits; the dimension of the 
state space $\cC$ corresponding to the protected DOF. 3bit QEC means 
the standard QEC code for weak phase noise \cite{NielsenBook}. Note 
that finite-distance QEC could also be used to protect one qubit 
against general weak noise (either independent or collective), but 
the smallest such code requires 5 qubits \cite{Knill-5BitQEC}. 
\label{NoiseComp} }
\end{center}
\end{table}

\newpage

\newpage

\begin{table}
\begin{center}
\begin{tabular}{|c|ccc|c|}
\hline
$\;\;${Quantum process}$\;\;\;\;$ & $\;\;\;F_{|z\rangle}\;\;\;$ 
& $\;\;\;F_{|x\rangle}\;\;\;$ 
& $\;\;\;F_{|y\rangle}\;\;\;$
& $\;\;\;\; {\bf F_e}\;\;\;\;$ \rule[-2mm]{0mm}{7mm}\\ \hline\hline
${\cal Q}_{x,{\sf un}}$	& 0.50 & 0.97 & 0.49 & {\bf 0.48} 
\\ \hline
${\cal Q}_{0,{\sf ns}}$	& 0.84 & 0.74 & 0.78 & {\sf 0.68} 
\\ 
${\cal Q}_{x,{\sf ns}}$	& 0.79 & 0.74 & 0.78 & {\sf 0.66} 
\\ 
${\cal Q}_{y,{\sf ns}}$	& 0.81 & 0.77 & 0.82 & {\sf 0.70} 
\\ 
${\cal Q}_{z,{\sf ns}}$	& 0.86 & 0.72 & 0.76 & {\sf 0.67} 
\\ \hline
$\;\;{\cal Q}_{zx,{\sf un}}$	& 0.49 & 0.50 & 0.50 & {\bf 0.24} 
\\ \hline
$\;\;{\cal Q}_{00,{\sf ns}}$	& 0.80 & 0.79 & 0.80 & {\sf 0.70} 
\\ 
$\;\;{\cal Q}_{zx,{\sf ns}}$	& 0.78 & 0.80 & 0.82 & {\sf 0.70} 
\\ 
$\;\;{\cal Q}_{zy,{\sf ns}}$	& 0.79 & 0.80 & 0.82 & {\sf 0.70} 
\\ 
$\;\;\;{\cal Q}_{000,{\sf ns}}$ & 0.77 & 0.79 & 0.78 & {\sf 0.67} 
\\ 
$\;\;\;{\cal Q}_{yzx,{\sf ns}}$ & 0.75 & 0.80 & 0.77 & {\sf 0.66} 
\\ 
\hline \hline
\end{tabular}
\end{center}
\caption{ Experimental data for implementation of various collective 
error models in the strong noise limit. The first column lists the 
one-bit quantum channels realized in the experiment.  
In addition to the applied error model, $ {\cal E}_x, {\cal E}_y, 
{\cal E}_z, {\cal E}_{zx}, {\cal E}_{zy}, {\cal E}_{yzx} $, the 
channel label specifies whether ({\sf ns}) or not ({\sf un}) encoding 
and decoding procedures were implemented.  The processes 
${\cal Q}_{0,{\sf ns}}$, ${\cal Q}_{00,{\sf ns}}$, 
${\cal Q}_{000,{\sf ns}}$ differ in the duration over which the net 
identity evolution is applied.  For each process, the input-output 
fidelities $F_{|\psi_{in}\rangle}$ involved in the process tomography 
as well as the resulting entanglement fidelities $F_e$ are listed. 
Statistical uncertainties are $\sim 2\%$, arising from errors in 
the tomographic density matrix reconstruction.  } 
\label{StrongNoiseData}
\end{table}

\begin{table}
\begin{center}
\begin{tabular}{rrcll}
\hspace{7pt} ${\cal Q}_{x,{\sf un}}$ & & $F_e = 0.48$ & & \\
$\openone$ & ~~~$\sigma_x$ & $\sigma_y$~~~ & \hspace{-4mm} $\sigma_z$& \\
				  & & & & $\openone$ \\
				  & & & & $\sigma_x$ \\
\multicolumn{4}{c}{
\raisebox{7.5pt}[0pt]{ 
\begin{math} 
\left(
\begin{tabular}{cccc}
~1.00 & ~0.01 & ~0.00 & -0.01~ \\
~0.00 & ~0.92 & ~0.06 & -0.19~ \\
~0.00 & -0.02 & -0.02 & ~0.02~ \\
~0.00 & ~0.07 & ~0.00 & ~0.02~ \\
\end{tabular} 
\right) 
\end{math} 
}
} & $\sigma_y$ \\
& & & & \hspace{-7pt} \raisebox{5mm}[0pt]{ $\:\sigma_z$} \\
\end{tabular}

\vspace*{3mm}

\begin{tabular}{rrcll}
\hspace{7pt} ${\cal Q}_{0,{\sf ns}}$ & & $F_e = 0.69$ & & \\
$\openone$ & ~~~$\sigma_x$ & $\sigma_y$~~~ & \hspace{-4mm} $\sigma_z$& \\
		& & & & $\openone$ \\
		& & & & $\sigma_x$ \\
\multicolumn{4}{c}{
\raisebox{7.5pt}[0pt]{ 
\begin{math} 
\left(
\begin{tabular}{cccc}
~1.00 & -0.03 & ~0.03 & ~0.02~ \\
~0.00 & ~0.53 & ~0.29 & -0.16~ \\
~0.00 & -0.25 & ~0.56 & ~0.05~ \\
~0.00 & ~0.10 & -0.07 & ~0.67~ \\
\end{tabular} 
\right) 
\end{math} 
}
} & $\sigma_y$ \\
& & & & \hspace{-7pt} \raisebox{5mm}[0pt]{ $\:\sigma_z$} \\
\end{tabular}

\vspace*{3mm}

\begin{tabular}{rrcll}
\hspace{7pt} ${\cal Q}_{x,{\sf ns}}$ & & $F_e = 0.67$ & & \\
$\openone$ & ~~~$\sigma_x$ & $\sigma_y$~~~ & \hspace{-4mm} $\sigma_z$& \\
		& & & & $\openone$ \\
		& & & & $\sigma_x$ \\
\multicolumn{4}{c}{
\raisebox{7.5pt}[0pt]{ 
\begin{math} 
\left(
\begin{tabular}{cccc}
~1.00 & -0.03 & ~0.02 & ~0.02~ \\
~0.00 & ~0.54 & ~0.28 & -0.21~ \\
~0.00 & -0.20 & ~0.56 & ~0.09~ \\
~0.00 & ~0.21 & -0.07 & ~0.58~ \\
\end{tabular} 
\right) 
\end{math} 
}
} & $\sigma_y$ \\
& & & & \hspace{-7pt} \raisebox{5mm}[0pt]{ $\:\sigma_z$} \\
\end{tabular}

\vspace*{3mm}

\begin{tabular}{rrcll}
\hspace{7pt} ${\cal Q}_{y,{\sf ns}}$ & & $F_e = 0.69$ & & \\
$\openone$ & ~~~$\sigma_x$ & $\sigma_y$~~~ & \hspace{-4mm} $\sigma_z$& \\
		& & & & $\openone$ \\
		& & & & $\sigma_x$ \\
\multicolumn{4}{c}{
\raisebox{7.5pt}[0pt]{ 
\begin{math} 
\left(
\begin{tabular}{cccc}
~1.00 & ~0.02 & ~0.05 & ~0.03~ \\
~0.00 & ~0.55 & ~0.27 & -0.16~ \\
~0.00 & -0.18 & ~0.61 & ~0.11~ \\
~0.00 & ~0.16 & -0.09 & ~0.60~ \\
\end{tabular} 
\right) 
\end{math} 
}
} & $\sigma_y$ \\
& & & & \hspace{-7pt} \raisebox{5mm}[0pt]{ $\:\sigma_z$} \\
\end{tabular}

\vspace*{3mm}

\begin{tabular}{rrcll}
\hspace{7pt} ${\cal Q}_{z,{\sf ns}}$ & & $F_e = 0.70$ & & \\
$\openone$ & ~~~$\sigma_x$ & $\sigma_y$~~~ & \hspace{-4mm} $\sigma_z$& \\
		& & & & $\openone$ \\
		& & & & $\sigma_x$ \\
\multicolumn{4}{c}{
\raisebox{7.5pt}[0pt]{ 
\begin{math} 
\left(
\begin{tabular}{cccc}
~1.00 & -0.10 & ~0.00 & ~0.03~ \\
~0.00 & ~0.56 & ~0.29 & -0.10~ \\
~0.00 & -0.19 & ~0.54 & ~0.03~ \\
~0.00 & ~0.12 & -0.07 & ~0.67~ \\
\end{tabular} 
\right) 
\end{math} 
}
} & $\sigma_y$ \\
& & & & \hspace{-7pt} \raisebox{5mm}[0pt]{ $\:\sigma_z$} \\
\end{tabular}

\vspace*{3mm}

\end{center}
\caption{ Experimentally reconstructed super-operators for the 
one-qubit channels.  A fixed row represents the decomposition along 
the set of Pauli operators $\{\openone, \sigma_x, \sigma_y, \sigma_z\}$ 
of the output corresponding to a given operator $\openone, \sigma_x, 
\sigma_y$, or $\sigma_z$ taken as input. 
Because trace-preservation is assumed, the first entry in each row
is constrained to unity, and the remaining entries in the first column 
to zero. Non-zero off-diagonal elements on the first row indicate the 
non-unitality of the channel.  
Entanglement fidelities, calculated via Eq. (\ref{SchumacherFidEnt})
are listed on top for each channel, and are also consistent with 
the values calculated from Eq. (\ref{AvgCorr}), 
which equals $\mbox{Tr}\{\cQ\}/4$. The above data may be interpreted 
as providing a matrix representation of the underlying super-operator
as in \cite{Havel-SuperOp} upon appropriately transposing rows and colums. }
\label{SuperOperator}
\end{table}

\end{document}